\documentclass[a4paper,12pt]{article}
\pdfoutput=1
\usepackage{jheppub}
\usepackage[T1]{fontenc}
\usepackage{cancel,tensor,enumitem,fix-cm}
\usepackage{epsfig}
\usepackage{graphicx}
\usepackage[english]{babel}
\usepackage{amsmath}
\usepackage{textcomp}
\usepackage{multirow}
\usepackage{booktabs}
\newcommand{\ra}[1]{\renewcommand{\arraystretch}{#1}}

\graphicspath{{Figures/}}

\subheader{\begin{flushright}
\end{flushright}}

\title{Fast and Slow Coherent Cascades in Anti-de Sitter Spacetime}

\author{Fotios V. Dimitrakopoulos,}
\author{Ben Freivogel}
\author{and Juan F. Pedraza}
\affiliation{GRAPPA and ITFA, Institute of Physics, Universiteit van Amsterdam,\\
Science Park 904, 1090 GL Amsterdam, Netherlands}
\emailAdd{f.dimitrakopoulos@uva.nl}
\emailAdd{benfreivogel@gmail.com}
\emailAdd{jpedraza@uva.nl}

\abstract{We study the phase and amplitude dynamics of small perturbations in 3+1 dimensional Anti-de Sitter spacetime using the truncated resonant approximation, also known as the Two Time Framework (TTF). We analyse the phase spectrum for different classes of initial data and find that higher frequency modes turn on with coherently aligned phases. Combining numerical and analytical results, we conjecture that there is a class of initial conditions that collapse in \emph{infinite slow time} and to which the well-studied case of the two--mode, equal energy initial data belongs. We additionally study perturbations that collapse in finite time, and find that the energy spectrum approaches a power law, with the energy per mode scaling approximately as the inverse first power of the frequency.}

\begin{document}

\maketitle

\flushbottom

\section{Introduction}

The question of stability of global Anti-de Sitter spacetime under small perturbations was energized by the seminal work of P. Bizon and A. Rostworowski \cite{BizRos11}, in which they traced the evolution of small perturbations in the form of a spherically symmetric scalar field. For most of the field profiles they studied, they found that the perturbation would eventually collapse to a black hole and the time of the collapse scales as the inverse square power of the amplitude $\epsilon$ of the perturbation. However, subsequent work on the subject has revealed a very rich and intriguing dynamics of the problem \cite{DiaHor11,DiaHor12,MalRos13a,BucLie13,AbadaS14,BalBuc14,CraEvn14,BasKri14,HorSan14,DFLY14,CraEvn14a,Yan15,BasKri15,Evnin:2015gma,HasNis15,MasSer15,DimYan15,GreMai15,JokPon15,DepFre16,CraEvn15a,CraEvn15b,DiaSan15,FreYan15,CarRoc16,DiaSan16,OliSop16,Deppe16,DFPY16,Ros16}.

Many initial data that do not collapse were found in a number of works \cite{BucLie13,GreMai15,MalRos13a,DepFre16,HorSan14,DiaSan16,Ros16}. At the same time, analytic studies suggested that these non-collapsing solutions may survive in the limit of vanishing amplitude \cite{DFLY14, DimYan15}. While numerical results are useful to study finite--energy perturbations, it is interesting to address this question in the zero--energy limit. Perturbation theory  \cite{BalBuc14, CraEvn14, DFLY14} has proven to be quite fruitful in this direction. Remarkably, the resulting system of equations at first non-linear order possesses a scaling symmetry which allows one to draw conclusions about the zero amplitude
($\epsilon \rightarrow 0$) limit from results at finite $\epsilon$.
Consequently, this scaling symmetry can be used to show that `stability islands' persist in the zero amplitude limit.\footnote{These regions of stability are not actually \emph{islands} since they are connected to each other through empty AdS  \cite{DimYan15}.}

In momentum space, the evolution of small perturbations towards a black hole translates into an efficient transfer of energy to modes with high frequencies. Given that the spectrum of linearized perturbations around AdS is commensurate, efficient energy transfer can occur and conventional perturbation theory may break down. Various modifications to the standard perturbation theory have been proposed, and are known under different names: \emph{two time scale framework} (TTF) \cite{BalBuc14}, \emph{renormalization group perturbation method} \cite{CraEvn14,CraEvn14a} or \emph{resonant approximation} \cite{BizMal15}.
All of these techniques are essentially equivalent and capture efficiently the transfer of energy between different modes up to the first nonlinear timescale $t\sim \epsilon ^{-2}$. This is accomplished by introducing a \emph{slow time} dependence of the amplitudes and the phases of the linearized modes.

In this paper we use the Two Time Framework (TTF) to study the evolution of the phases and amplitudes of the Fourier modes, for a variety of initial conditions. In Sec.(\ref{sec:Review}) we present a brief review of the formalism. Quite generally, we find that the phases are not excited in a random way, as in the standard theory of weak turbulence, but in an almost coherent way. This supports the coherent phase conjecture proposed in \cite{FreYan15}. However, we observe small deviations from a strictly coherent spectrum at the moment of collapse, suggesting that the proposal of \cite{FreYan15} could be refined.

In Sec.(\ref{sec:2ModeData}) we study the evolution of initial data where only the first two modes are initially turned on. We give particular attention to the case of equal energy initial data. Using a combination of numerical and analytic results, we conjecture that they belong to a class of initial conditions that collapse in {\bf infinite slow time}.\footnote{In the limit of vanishing amplitude, $\epsilon \rightarrow 0.$} A nice way to diagnose singularity formation in the numerical evolutions is through the so-called \emph{analyticity strip method} \cite{AStrip83,BizJal13,BizMal15}. This method can be used to estimate the time at which the amplitude spectrum becomes power-law at large mode numbers, thus signaling a large backreaction. Our numerical results suggest that that, for the two-mode equal-energy data, the width of the analyticity strip goes to zero {\bf exponentially} in time.

Moreover, we show analytically that solutions dominated by the two lowest two modes obey a simple speed limit on the rate of energy transfer to higher frequencies. The speed limit depends on the frequency and the number of dimensions. As a result, we show that in $3+1$ dimensions it takes infinite slow time to transfer energy to arbitrarily high frequency, while in higher dimensions it takes finite slow time.\footnote{This statement \emph{does not} exclude the existence of direct (or inverse) cascades within TTF.}  Because our analytic argument assumes the system is dominated by the lowest two modes, it cannot be used to rigorously infer the late-time behavior of the system; however, it is intriguing that our simple argument agrees with numerics.

There has been a long-lasting dispute about the ultimate fate of the two mode, equal-energy data \cite{BalBuc14,BizRos14,DepFre16}.\footnote{The fact that this data might collapse in infinite slow time was suggested earlier to us by Andrzej Rostworowski, as we understand, based on simulations of full General Relativity in AdS. It is our understanding that A. Rostworowski and collaborators have also obtained interesting results about this $3+1$-dimensional problem, and we look forward to comparing them to our results.} Our results suggest that, for this initial condition, the time scale required to collapse is slightly larger than the standard one, e.g. $t\sim\epsilon^{-2}\log\epsilon$ instead of $t\sim\epsilon^{-2}$, which is not fully captured within TTF. It would be interesting to extend the TTF formalism to include up to the second non-linear time scale and repeat the analysis performed in this paper.

In Sec.(\ref{sec:GaussianData}) we study the numerical evolution of initial data with Gaussian profile. As expected, we find that the width of the analyticity strip goes to zero in finite slow time, as long as the wavepacket is sufficiently narrow. However, at the time of collapse the amplitudes $A_n$ of the normal modes approach a power-law spectrum of the form $A_n \sim n^{-\gamma}$ with $\gamma \approx 3/2$, which slightly modifies the result $\gamma = 8/5$ reported in \cite{MalRos13}.\footnote{For $\gamma=3/2$, the total energy is logarithmically divergent in the ultraviolet. This leads to a weak dependence of the overall amplitude on the UV cutoff, but we are not aware of any physical argument against it. The fact that the exponent may be smaller than 8/5 was also conveyed to us in a discussion with A. Rostworowski and collaborators.} This amplitude spectrum corresponds to an energy spectrum of the form $E_n \sim 1/\omega_n\sim 1/n$. Interestingly, the form of this spectrum coincides with the one predicted in \cite{FreYan15} using a coherent phase ansatz.

\section{Spherically Symmetric Perturbations in AdS}\label{sec:Review}

For completeness, we will start in this section by reviewing the model of scalar field perturbations of \cite{BizRos11} as well as the TTF system \cite{BalBuc14, CraEvn14} and the \emph{phase coherent} conjecture of \cite{FreYan15}.

We consider a minimally coupled, spherically symmetric scalar field $\phi$ in asymptotically $AdS_4$. The ansatz for the metric is the following:
\begin{equation}
ds^2 = \frac{1}{\text{cos}^2x}\left( -A e^{-2\delta} dt^2 + A^{-1}dx^2 + \text{sin}^2x d\Omega ^2 \right),
\end{equation}
where, for simplicity, we have set the AdS radius to unity. Due to spherical symmetry $A$, $\delta$ and $\phi$ are only functions of time $t$ and the radial coordinate $x \in [0, \frac{\pi}{2}) $. It is customary to introduce the auxiliary variables $\Phi = \phi ' $ and $\Pi = A^{-1}e^{\delta} \dot{\phi}$. In terms of these variables, the equation of motion for $\phi$ can be written as:
\begin{eqnarray}\label{eq:EOM1}
\dot{\Phi} =  \left( A e^{- \delta} \Pi \right) ' , \qquad \dot{\Pi}  =   \frac{1}{\text{tan}^2x} \left( \text{tan}^2 x A e^{-\delta} \Phi \right) ' ,
\end{eqnarray}
while the Einstein equations reduce to the constraints:
\begin{eqnarray}\label{eq:EOM2}
A' & = & \frac{1 + 2\, \text{sin}^2x}{\text{sin}x\,\text{cos}x} \left( 1- A \right) - \text{sin}x\, \text{cos}x\, A \left( \Phi ^2 + \Pi ^2 \right) \nonumber \\
\delta ' & = & - \text{sin}x\, \text{cos}x \left( \Phi ^2 + \Pi ^2 \right).
\end{eqnarray}
The overdots and the primes here denote differentiation with respect to $t$ and $x$ respectively. One can try and solve this system analytically in perturbation theory, expanding the fields around the pure AdS solution: $\phi = 0$, $A=1$ and $\delta = 0$. Assuming that the initial perturbation is of order $\phi(t=0)\sim\epsilon$ we can write the following series expansions:
\begin{eqnarray}
\phi & = & \sum_{j=0}^{\infty} \epsilon ^{2j+1} \phi_{2j+1}(t,x), \quad A = 1-\sum_{j=0}^{\infty}\epsilon ^{2j} A_{2j}(t,x), \quad \delta = \sum_{j=0}^{\infty}\epsilon ^{2j} \delta_{2j}(t,x).
\end{eqnarray}
Inserting this ansatz into the equations of motion and collecting terms at each order in $\epsilon$ we obtain a set of linear equations which can be solved order by order. To first order, we obtain the equation of scalar filed propagating in empty AdS,
\begin{eqnarray}
\ddot{\phi}_{1} + L \phi _1 = 0\,.
\end{eqnarray}
Here, $L = - \frac{1}{\text{tan}^2 x} \partial _x \left( \text{tan}^2 x \partial _x \right)$ is the Laplacian operator in AdS. The eigenvalues of this operator are $\omega _j = (2j +3)$, for $j\in\{0,1,2,\ldots\}$; the fact that they are positive definite implies that AdS is linearly stable. The eigenfunctions are
\begin{eqnarray}
e_j = d_j\, \text{cos}^3 x\, _2F_1\left( -j,3+j,3/2 ; \text{sin}^2 x \right), \qquad d_j =\sqrt{\frac{16}{\pi}(j+1)(j+2)},
\end{eqnarray}
These fuctions are normalized such that $\left(e_m , e_n \right) = \delta _{mn}$ where the inner product is defined as $\left(f,g \right)\equiv \int_{0}^{2 \pi} f(x) g(x) \text{tan}^2 x\, dx$. One can use the eigenfunctions of $L$ as a basis for $\phi$. At the leading order in $\epsilon$ we have:
\begin{equation}
\phi _1 (t,x) = \sum_{j=0}^{\infty} c^{(1)}_j (t) e_j(x) = \sum_{j=0}^{\infty} \left( \alpha _j e^{i \omega _j t} + \bar{\alpha}_j e^{-i \omega _j t}  \right) e_j(x) .
\end{equation}
To second order in $\epsilon$ we obtain the backreaction on the metric:
\begin{eqnarray}
A_2(t,x) & = & -\nu (x) \int _{0}^{x} \left( \dot{\phi} _1 (t,y) ^2 + \phi_1 ^{\prime} (t,y) ^2 \right) \mu (y) dy, \label{eq:A2}\\
\delta _2 (t,x) & = & \begin{cases}
                                 - \int _{0}^x  \left( \dot{\phi} _1 (t,y) ^2 + \phi_1 ^{\prime} (t,y) ^2 \right) \nu (y) \mu (y) dy, \quad \text{for} \; \delta (t,0) = 0,\\
                                  \int _{x}^{\pi /2}  \left( \dot{\phi} _1 (t,y) ^2 + \phi_1 ^{\prime} (t,y) ^2 \right) \nu (y) \mu (y) dy, \quad \text{for} \; \delta (t,\pi / 2) = 0,\label{eq:delta2} \\
                                 \end{cases}
\end{eqnarray}
where
\begin{equation}
\mu(x)=\tan^2x\,,\qquad\nu(x)=\frac{\sin x \cos x}{\mu(x)}\,.
\end{equation}
The two choices in (\ref{eq:delta2}) correspond to different time gauges, which are physically equivalent \cite{DFPY16}.
The first non--trivial dynamics of the scalar field appear at the third order in $\epsilon$. At this order one obtains the following inhomogeneous equation
\begin{eqnarray}
\ddot{\phi}_3 + L \phi _3 = S\left(\phi _1, A_2, \delta _2 \right),
\end{eqnarray}
where $ S = -2 \ddot{\phi}_1 \left(A_2 + \delta _2  \right) - \dot{\phi}_1 \left( \dot{A}_2 + \dot{\delta}_2 \right) - \phi _1 ^{\prime} \left( A_2^{\prime} + \delta _2 ^{\prime} \right)$.
We can again expand the field as $\phi _{3}(t,x) = \sum _{j=0}^{\infty} c^{(3)}_j (t) e_j (x)$. Projecting onto the basis $\{ e_m \}$, we obtain an infinite set of decoupled forced harmonic oscillators for the Fourier coefficients:
\begin{eqnarray}
\ddot{c}^{(3)}_j (t) + \omega _j ^2 c^{(3)}_j (t) = S_j, \quad S_j \equiv \left( S , e_j \right).
\end{eqnarray}
Due to the fact that the spectrum is highly commensurate, there can be many resonant terms contained in $S_j$. Specifically, for every triad\footnote{$S_j$ is cubic in $a_j$  at this order.} $\left( j_1 , j_2 , j_3 \right)$ such that $\omega _j = \omega _{j_1} + \omega _{j_2} - \omega _{j_3}$ there is a resonant term in $S_j$. A first analysis of these resonances appeared in \cite{BizRos11} but a more sophisticated formalism, the Two Time Framework, was later developed in \cite{BalBuc14,CraEvn14} allowing for a more systematic treatment. The basic idea behind TTF is to allow for an additional {\it slow time} dependence of the fields. In that case, the perturbative expansion of the fields becomes
\begin{eqnarray}\label{eq:slowtimeexpansion}
\phi & = & \epsilon \phi _1 (t,\tau , x) + \epsilon ^3 \phi _3 (t, \tau, x) + \mathcal{O}(\epsilon ^5), \nonumber \\
A & = & 1 - \epsilon ^2 A_2 (t,\tau , x) + \mathcal{O}(\epsilon ^4), \nonumber \\
\delta _2 & = & \epsilon ^2 \delta _2 (t,\tau , x) + \mathcal{O}(\epsilon ^4),
\end{eqnarray}
where $\tau = \epsilon ^2t$ denotes the slow time. One could go to higher orders by introducing additional slow time variables. Substituting the above expansion, eq.~(\ref{eq:slowtimeexpansion}), into the equations of motion, eqs.~(\ref{eq:EOM1})-(\ref{eq:EOM2}), one obtains a system of equations which can be solved order by order in $\epsilon$. The difference is that the expansion for the scalar field $\phi_1$ (at leading order in $\epsilon$) admits now solutions of the form
\begin{eqnarray}\label{eq:phi1inTTF}
\phi _{1}(t,\tau, x) = \sum_{j=0}^{\infty} \left( \alpha _j (\tau) e^{-i \omega _j t} + \bar{\alpha} _j (\tau) e^{-i \omega _j t} \right) e_j (x)\,.
\end{eqnarray}
The rapid time $t$ accounts for the oscillations of the normal modes, while the slow time $\tau$ accounts for the energy transfer among the normal modes (due to non--linearities). To second order in $\epsilon$ we have again the constraints for $A_2$ and $\delta _2$ which can be easily integrated as in eqs.~(\ref{eq:A2})-(\ref{eq:delta2}). At third order in $\epsilon$ we obtain
\begin{eqnarray}\label{eq:TTF2}
\partial _t \phi _{3} + L \phi _3 + 2 \partial _t \partial _{\tau} \phi _1 = S(\phi _1 , A_2, \delta _2),
\end{eqnarray}
where $ S(\phi _1 , A_2, \delta _2) = (\dot{A}_2 - \dot{\delta} _2 )\dot{\phi}_1 -2 (A_2 - \delta _2)L \phi _1 + (A_2 ^{\prime} - \delta _2 ^{\prime}) \phi _1 ^{\prime} $. The overdots here represent derivatives with respect to the fast time $t$, as usual. Projecting again onto the basis $\{ e_m \}$, and substituting eq.~(\ref{eq:phi1inTTF}) we obtain
\begin{eqnarray}
\left( e_j , \partial _t ^2 \phi _3 + \omega _j ^2 \phi _3 \right) - 2 i \omega _j \left( \partial _{\tau} \alpha _j (\tau) e^{-i \omega _j t} - \partial _{\tau} \bar{\alpha}_j (\tau) e^{i \omega _j t}  \right) = \left( e_j , S \right).
\end{eqnarray}
We can treat the resonances separately, due to the presence of terms proportional to $e^{\pm i \omega _j t}$ on the left hand side. We may cancel off these resonances from the rest of the eq.~(\ref{eq:TTF2}) by setting
\begin{eqnarray}\label{eq:TTF}
-2 i \omega _j \partial _{\tau}\alpha _j (\tau) = \left( e_j , S \right) [ \omega _j ] = \sum _{klm}S_{jklm} \bar{\alpha}_k \alpha _l \alpha _m\,.
\end{eqnarray}
The resonances are now entirely captured by eq.~(\ref{eq:TTF}), rendering the solution for $\phi _3$ bounded and hence, of little interest to us. The important result here is
given by eq.~(\ref{eq:TTF}), an infinite set of coupled first order ODEs for the Fourier coefficients of $\phi_1$, which are known as the {\it TTF equations}. The sum in eq.~(\ref{eq:TTF}) contains only terms for which the {\it resonance condition} $j + k =   l + m$ is satisfied.\footnote{In principle, all the resonant {\it channels}, $j = \pm k \pm j \pm l$, should appear but only the one described here are nonzero \cite{CraEvn14, Yan15}.} The TTF equations govern the evolution of both the amplitudes and the phases of the complex coefficients $\alpha_j(\tau)$. Hence, a more convenient way to write them, is by going to the {\it amplitude--phase representation} $\alpha _j (\tau) = A_{j}(\tau) e^{i B_j (\tau)} $, in terms of which the TTF equations can be expressed as
\begin{eqnarray}
2 \omega _{j} \frac{d A_j}{d \tau} & = & \sum_{\substack{j+k = l+m \\ j \neq l, j\neq m}} S_{jklm} A_k A_l A_m \ \text{sin} \left( B_j + B_k - B_l - B_m \right)\label{eq:amplitudes} \\
2 \omega _j \frac{d B_j}{d \tau} & = & T_j A_j ^2 + \sum_{i \neq j} R_{ij}A_i ^2 + A_j ^{-1} \sum_{\substack{j+k = l+m \\ j \neq l, j\neq m}} S_{jklm} A_k A_l A_m \ \text{cos} \left( B_j + B_k - B_l - B_m \right). \label{eq:phases} \nonumber \\
\end{eqnarray}
Here $T_j = S_{jjjj}$, $R_{ji} = S_{ijji} + S_{jiji}$, for $i \neq j$.  This system of equations is invariant under $\alpha _j (\tau) \rightarrow \epsilon \alpha_j (\tau/\epsilon ^2)$. This means that the evolution of a perturbation of amplitude $A_j$ for a time $\tau$, will be the same as the evolution of the perturbation of amplitude $\epsilon A_j$, but for a longer time $\tau / \epsilon ^2$. As mentioned earlier, this important symmetry allows one to draw conclusions for the vanishing amplitude limit, from results obtained at finite $\epsilon$ (as long as the TTF approximations are valid \cite{DimYan15}).

\subsection{Phase-coherent power laws}

In \cite{FreYan15} it was suggested that a power-law spectrum alone is not sufficient for a black hole to form but, in addition, the phases need to be coherent at the moment of collapse. The coherence of phases is defined as a phase spectrum that is \emph{asymptotically} linear in the mode number $n$,
\begin{equation}\label{eq:phasecoherence}
B_n (\tau) = \gamma (\tau) n + \delta (\tau) + \dots,
\end{equation}
where the ellipsis represent subleading terms in $n$. Part of this work is devoted to verify numerically the degree of coherence of the phases for different initial conditions, and look for possible deviations of perfect coherence in different moments of the evolution. In addition, as it was already argued in \cite{FreYan15}, the case of $AdS_{3+1}$ is subtle and a numerical study of the evolution of the phases, using the actual interaction coefficients $S_{ijkl}$ (and not just their asymptotic values) is important.


\section{The two--mode data}\label{sec:2ModeData}

We will start by looking at two mode--initial data, which is parametrized as follows:
\begin{eqnarray}\label{eq-An2M}
A_n = \frac{\epsilon}{3} \left( \delta ^{0}_n + \kappa \delta ^{1}_n \right).
\end{eqnarray}
This is the case where we initially excite the lowest two modes with no relative phase. Here, $\epsilon$ denotes the amplitude of the perturbation, while $\kappa$ parametrizes the amount of energy in the second lowest mode. For concreteness, we have considered data with $\epsilon=1$ and different values of $\kappa$. However, as discussed before, the TTF equations are invariant under the scaling symmetry $\alpha_n (\tau) \rightarrow \epsilon \alpha_n (\tau/\epsilon^2)$. Therefore, the solution for different values of $\epsilon$ would be completely equivalent to the case $\epsilon=1$ but evolved to a different time $\tau$.

\subsection{Numerical results}

\noindent \underline{{\bf Equal energy:}}\\

Setting $\kappa = 3/5$ corresponds to a situation where the energy is equally distributed between the two modes. This case is very interesting since it has been argued to be in the borderline of one of the stability islands \cite{BalBuc14,BizRos14,DepFre16}. We will study the interplay between the value of $\Pi$ at the origin,\footnote{The value of $\Pi (t,0)$ corresponds to the Ricci scalar at the origin and therefore is expected to be a good indicator of singularity developement.} which remains finite in the resonant approximation, and the spectrum of the amplitudes and the phases, hoping to shed some light from a different perspective and clarify the fate of this initial condition.

In order to detect the formation of singularities from the spectrum we use the so-called analyticity strip method, introduced in \cite{AStrip83}, and employed for the first time in the context of the AdS instability problem in \cite{BizJal13}. The idea here is to consider the the analytic extension of $\phi(t,z)$ into the complex plane of the radial
variable $r\to z\in\mathbb{C}$. The function $\phi(t,z)$ will typically have complex singularities moving in time; if one of these singularities
hits the real axis, $\phi(t,r)$ becomes singular. The pair of singularities closest to the real axis are denoted as $z=x\pm i\rho$, so that $\rho$ determines the width of the analyticity strip around the real axis. Thus, if $\rho$ vanishes at some point during the evolution then $\phi(t,r)$ will be singular. Now, $\rho$ is encoded in the exponential decay of the Fourier coefficients $A_n\sim e^{-\rho n}$ (at large $n$), so it can be  obtained  from  the asymptotics  of a given numerical solution.

 \begin{figure}[tb]
\begin{center}
\includegraphics[width=7.25cm]{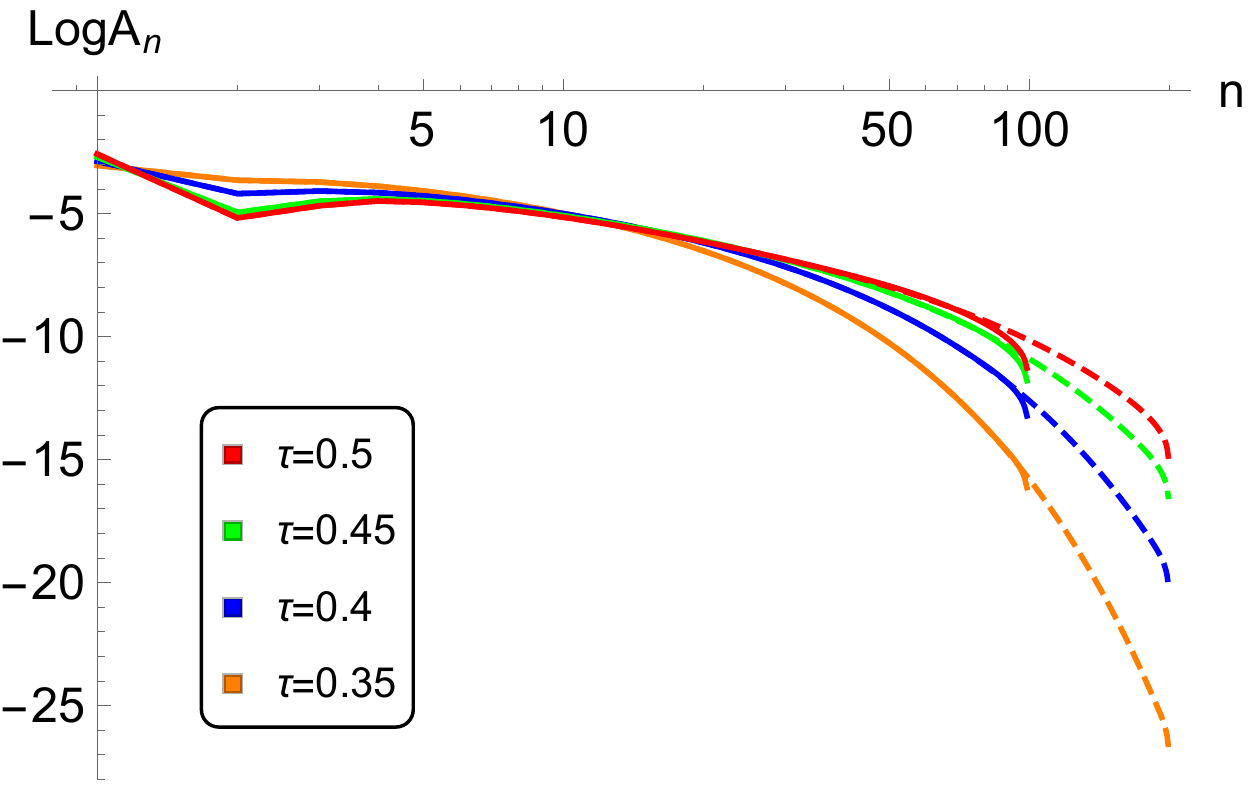}
\includegraphics[width=7.55cm]{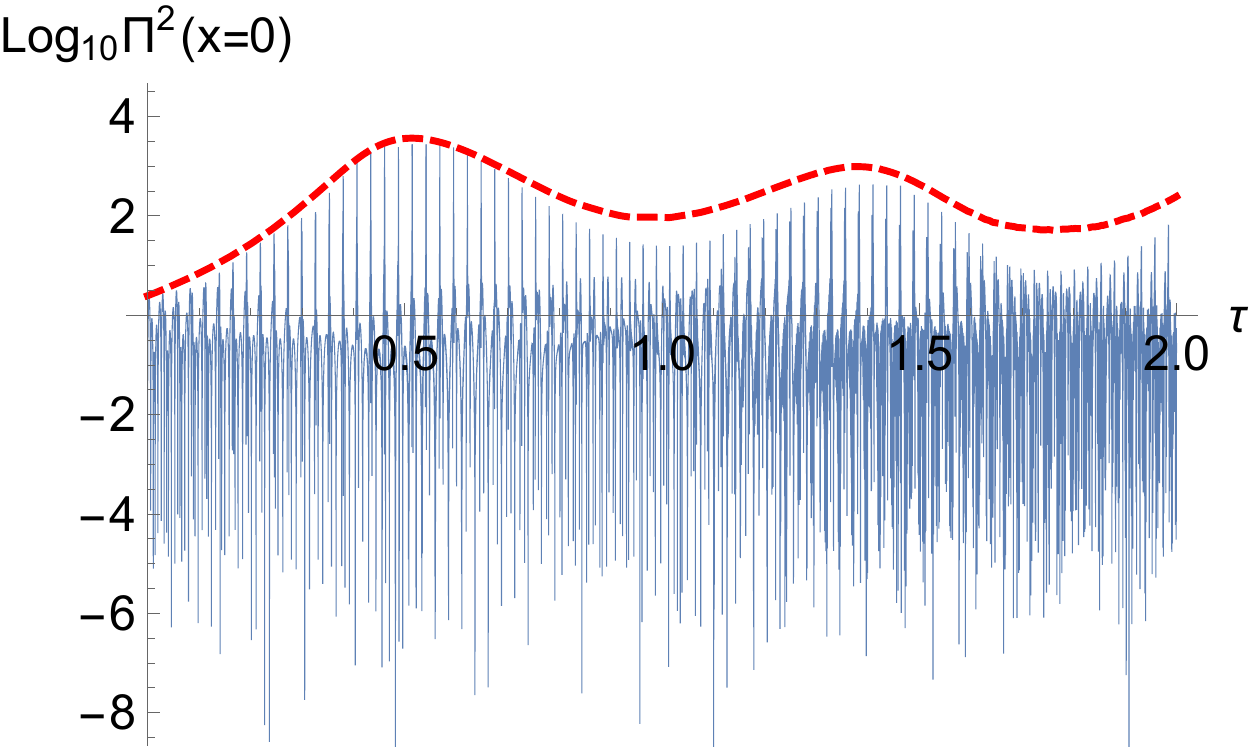}
\caption{{\bf Left:} Amplitude spectrum $A_n$, for different values of $\tau=\{0.35,5,0.45,5\}$. Beyond $\tau=0.5$ the spectrum develops small oscillations that are difficult to distinguish to the naked eye. Solid lines correspond to a 100 mode truncation, while dashed lines correspond to a 200 mode truncation. {\bf Right:} Ricci scalar at the origin, for a 200 mode truncation and up to $\tau=2$. For this plot we have chosen $\epsilon=0.09$, so we could directly compare with \cite{BalBuc14}. The red curve represents an estimate for the upper envelope assuming perfect phase alignment.
\label{fig-AnEqualE}}
\end{center}
\end{figure}

Before evolving the system (\ref{eq:amplitudes})-(\ref{eq:phases}) we must fix the gauge and find the coefficients $S_{ijkl}$ up to a maximum number $i=j=k=l=n_{\text{max}}$. We choose to work in the boundary gauge, where $\delta(t,\pi/2)=0$, since in this gauge the evolution of the phases is regular \cite{DFPY16,CraEvn15a,Deppe16} and therefore, the numerical integration is easier. In order to estimate the effect of the cutoff $n_{\text{max}}$ we evolve the system for two different cases, first for $n_{\text{max}}=99$ and then for $n_{\text{max}}=199$. In Fig.~(\ref{fig-AnEqualE}) we show the amplitude spectrum in the two cases, for different values of $\tau$. In general they agree over a wide range of $n$, but differ mildly in the range $n\in(79,99)$. For this plot, we have only shown values of $\tau$ up to $\tau=0.5$ (approximately up to this time the energy flows to higher modes monotonically, via a direct cascade), even though we evolved the system up to $\tau=2$. After this value of $\tau$ the spectrum starts to mildly oscillate (energy flow in both ways, via direct and inverse cascades) but it is difficult to distinguish the variations with the naked eye. In the right panel of the same figure, we plot the evolution of the Ricci scalar at the center, $\Pi (x=0)$, up to the maximum time of evolution $\tau=2$ and for a 200 mode truncation. For this plot we have rescaled the solution such that $\epsilon=0.09$ (using the symmetry $\alpha _j (\tau) \rightarrow \epsilon \alpha_j (\tau/\epsilon ^2)$), so we could directly compare with the results of \cite{BalBuc14}, and found perfect agreement (two and al half bounces up to $\tau=2$). The interesting feature of this plot is the upper envelope, resulting from joining all maxima of the fast oscillations. The exact values of these maxima \emph{depend} on the choice of $\epsilon$, and are tedious to find numerically. However, it can be shown that the envelope satisfies
\begin{equation}
\Pi (x=0)\leq\left(2\sum_{n=0}^{n_{\text{max}}}\omega_n e_n(0)|\alpha_n(\tau)|\right)^2\,,
\end{equation}
which is independent of $\epsilon$. This bound is obtained assuming perfect phase alignment, i.e. that during each period of the lowest mode ($\Delta t=2\pi/\omega_0$), there is a time when all modes are simultaneously peaked. We also plot such a bound in Fig.~\ref{fig-AnEqualE} (right panel), depicted in red. Thus, comparing the upper envelope of the fast oscillations with the red curve can give us information about phase alignment. Interestingly, upon a simple inspection we can see that the phases start to decohere when the inverse cascade kicks off, and tend to cohere again as the Ricci scalar increases (energy is stored in higher modes), i.e. when the direct cascade dominates. This observation gives support to the coherent phase conjecture of \cite{FreYan15}.

\begin{figure}[tb]
\begin{center}
\includegraphics[width=7.4cm]{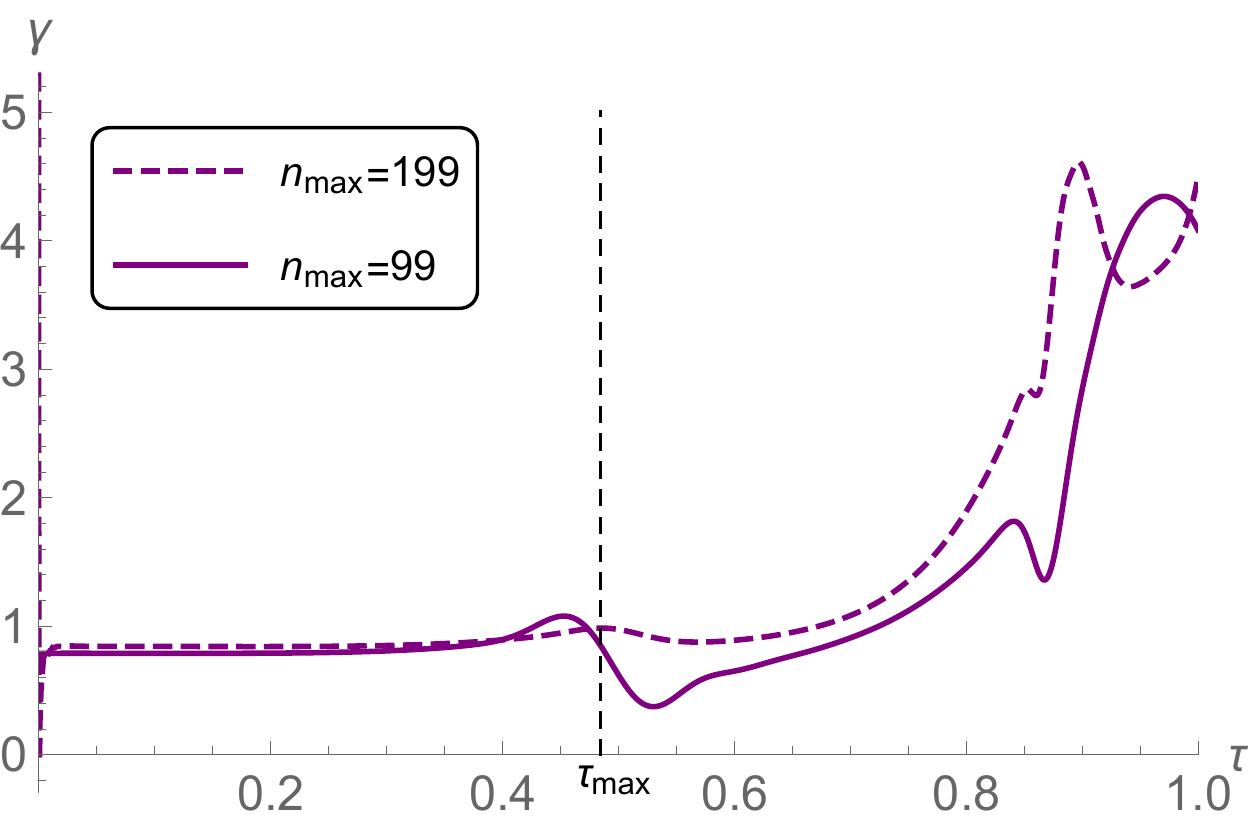}
\caption{Evolution of $\gamma(\tau)$ for $n_{\text{max}}=99$ and $n_{\text{max}}=199$. Up to $\tau=\tau_{\text{max}}$ the two curves are relatively equal to each other and almost constant in time, indicating that the dynamics up to this time is mostly encoded in $\rho(\tau)$. Beyond this point the two solutions differ and $\gamma(\tau)$ behaves quite erratically. It seems impossible to extrapolate $\gamma$ for $\tau\to\infty$.\label{gammafit}}
\end{center}
\end{figure}

It is interesting to
ask what happens with analyticity strip throughout the evolution, and how it correlates with the bounces observed in $\Pi(x=0)$. To answer this question, we fit the amplitudes using the following ansatz
\begin{eqnarray}\label{eq:ASansatz}
A_n(\tau) \sim \alpha (\tau)n^{-\gamma(\tau)}e^{-\rho(\tau)n}\,.
\end{eqnarray}
We exclude the modes $n\in(70,99)$ for $n_{\text{max}}=99$ and $n\in(140,199)$ for $n_{\text{max}}=199$ to avoid any cutoff effect. Also, since
the analyticity strip method applies only asymptotically, we also exclude the first modes, $n\in(0,29)$ for $n_{\text{max}}=99$ and $n\in(0,59)$ for $n_{\text{max}}=199$.
Comparing the two fittings should then provide a good test of our numerics.
\begin{figure}[tb]
\begin{center}
\includegraphics[width=7.4cm]{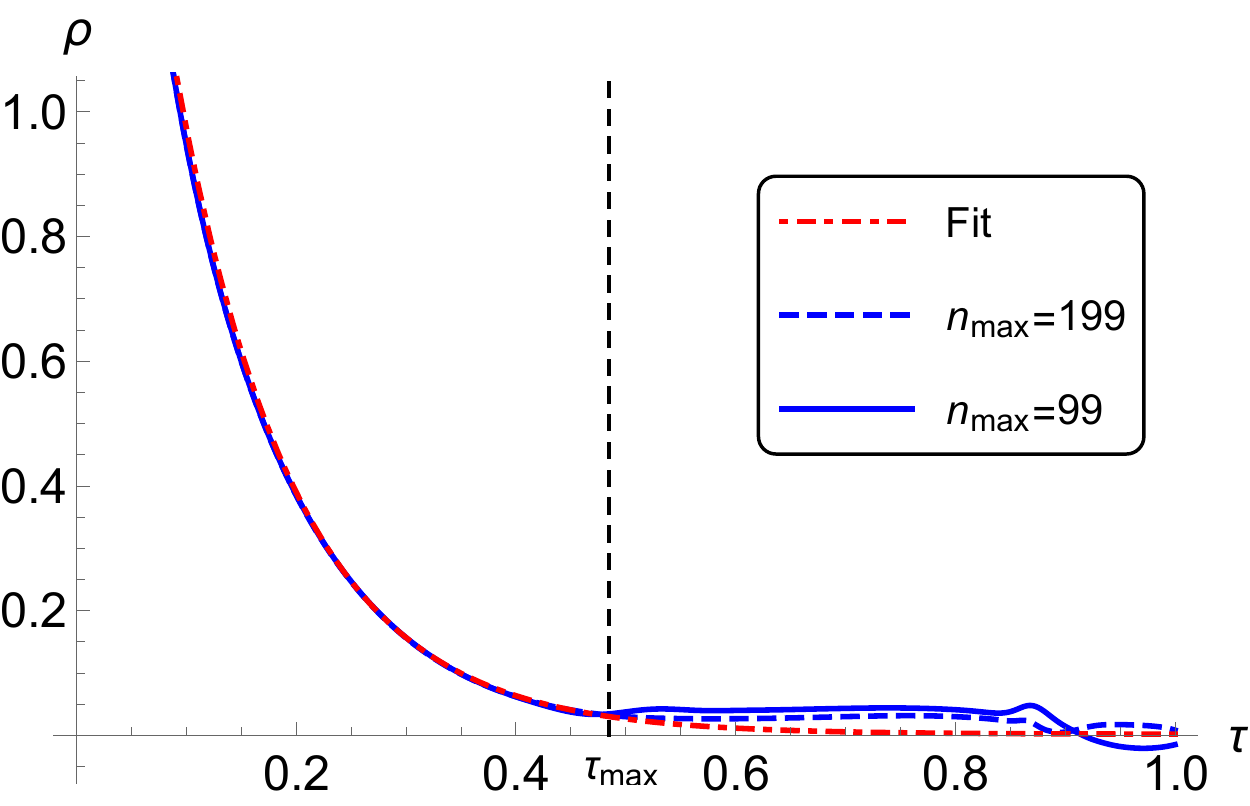}
\includegraphics[width=7.4cm]{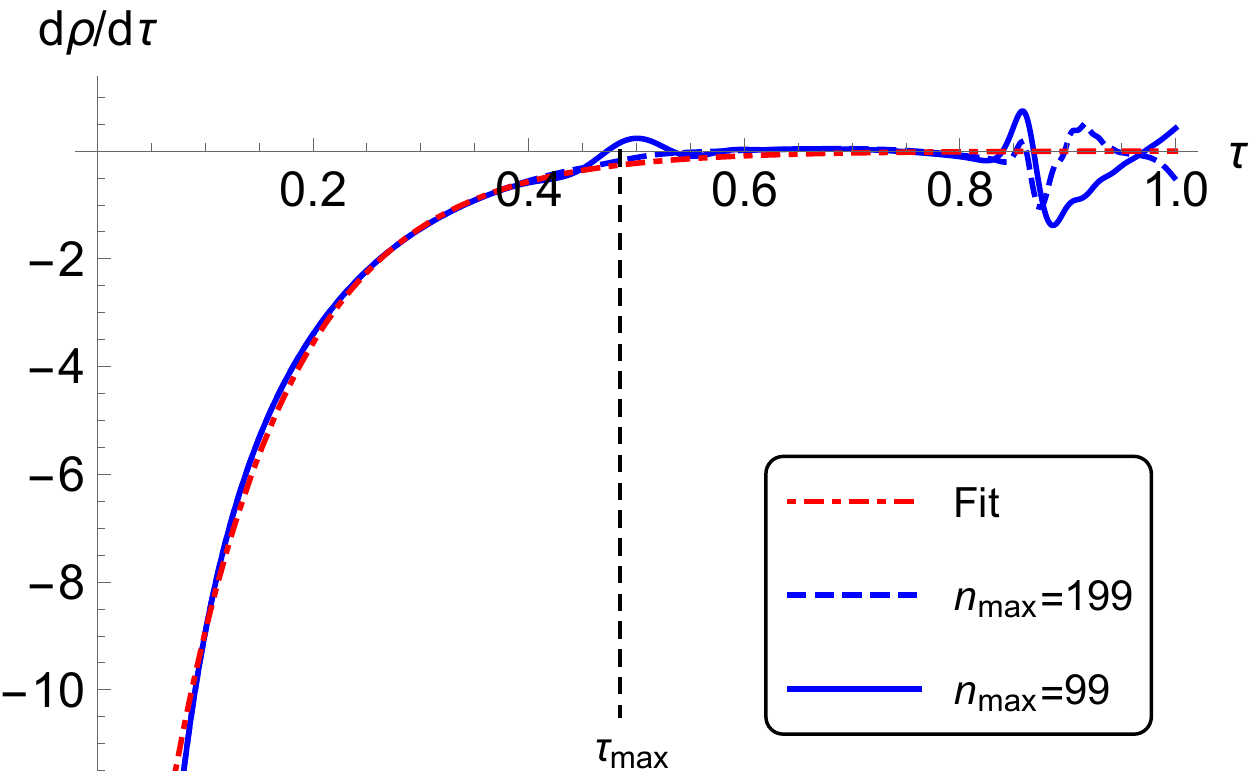}
\caption{ {\bf Left:} Evolution of $\rho(\tau)$ for $n_{\text{max}}=99$ and $n_{\text{max}}=199$. We define $\tau_{\text{max}}\sim0.485$ as the maximum time
such that $|\rho_{199}-\rho_{99}|\leq1/n_{\text{max}}$. We have also included the best fit up to this time, using the ansatz $\rho(\tau)=\rho_0e^{-\alpha \tau}+\rho_{\infty}$. This yields $\rho_{\infty}\sim5\times10^{-5}\ll1/n_{\text{max}}$, supporting the idea that $\rho\to0$ as $\tau\to\infty$.
{\bf Right:} Evolution of $d\rho/d\tau$ for $n_{\text{max}}=99$ and $n_{\text{max}}=199$. The behavior of the derivative for $\tau<\tau_{\text{max}}$ shows that the exponential function is indeed a good fit for $\rho$.
\label{rhoEE}}
\end{center}
\end{figure}
In Fig.~(\ref{rhoEE}) we plot the evolution of $\rho$ and its derivative as a function of $\tau$ for the two cases considered, $n_{\text{max}}=99$ and $n_{\text{max}}=199$. In general, they are in excellent agreement for the first part of the evolution, up to a $\tau_{\text{max}}$. On general grounds, we expect $\rho$ to be the most robust parameter in the fit since it appears in the exponential. Therefore, we define $\tau_{\text{max}}\sim0.485$ as the maximum time such that $|\rho_{199}-\rho_{99}|\leq1/n_{\text{max}}$. Beyond this point the curves start to differ, while staying \emph{exponentially small}.\footnote{We do not trust our results for the $n_\text{max}=99$ truncation beyond $\tau=\tau_{\text{max}}$. The solution for the $n_\text{max}=199$ truncation is expected to be valid for longer times, however, in order to determine a similar upper bound we would require a similar comparison with a solution obtained with a higher order truncation.} The fact $\rho$ stays small after $\tau=\tau_{\text{max}}$ suggests that the bounces observed in $\Pi(x=0)$ are more likely to be related to the other parameters of the fits and/or the phases, and not to $\rho$ itself. Also, notice that up to $\tau_{\text{max}}\sim0.485$, $\rho$ decays almost exponentially, so we can try to fit it as
\begin{eqnarray}
\rho(\tau)\sim \rho_0 e^{-\alpha \tau}+\rho_{\infty}\,.
\end{eqnarray}
Interestingly, we find that $\rho_{\infty}\sim5\times10^{-5}\ll1/n_{\text{max}}$ suggesting that, if this extrapolation can be trusted for longer times, $\rho\to0$ as $\tau\to\infty$. In Sec.(\ref{sec:speed}) we give an analytic argument to explain why this can happen in $(3+1)-$dimensions. The value of $\gamma$ is almost constant for $\tau<\tau_{\text{max}}$ ---see Fig.~(\ref{gammafit})--- but fluctuates after this time. This makes it impossible to extrapolate its value to $\tau\to\infty$. The behavior of $\gamma(\tau)$ should be contrasted with that of $\rho(\tau)$ which, on the contrary, stays small after $\tau=\tau_{\text{max}}$. This suggests that $\gamma(\tau)$ might have a more relevant effect on $\Pi(x=0)$ after $\tau=\tau_{\text{max}}$, and hence on the oscillations.

\begin{figure}[tb]
\begin{center}
\includegraphics[width=7.4cm]{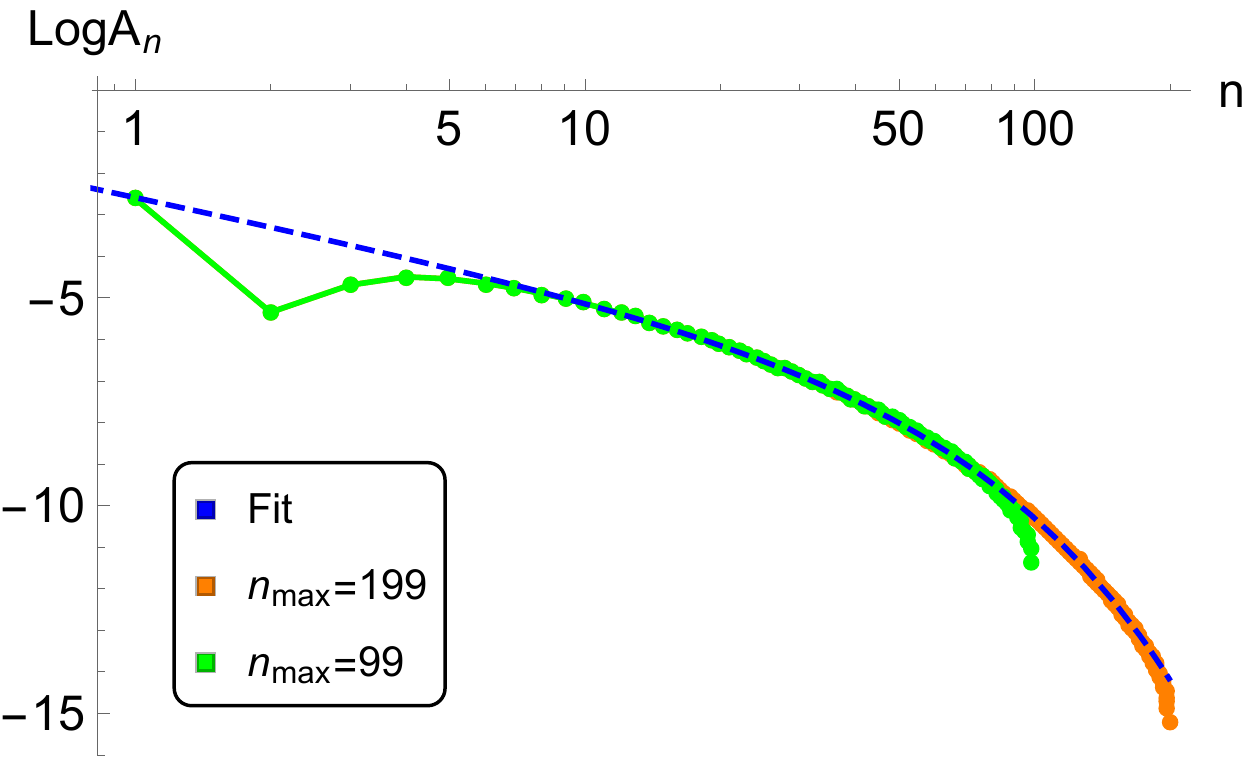}
\includegraphics[width=7.4cm]{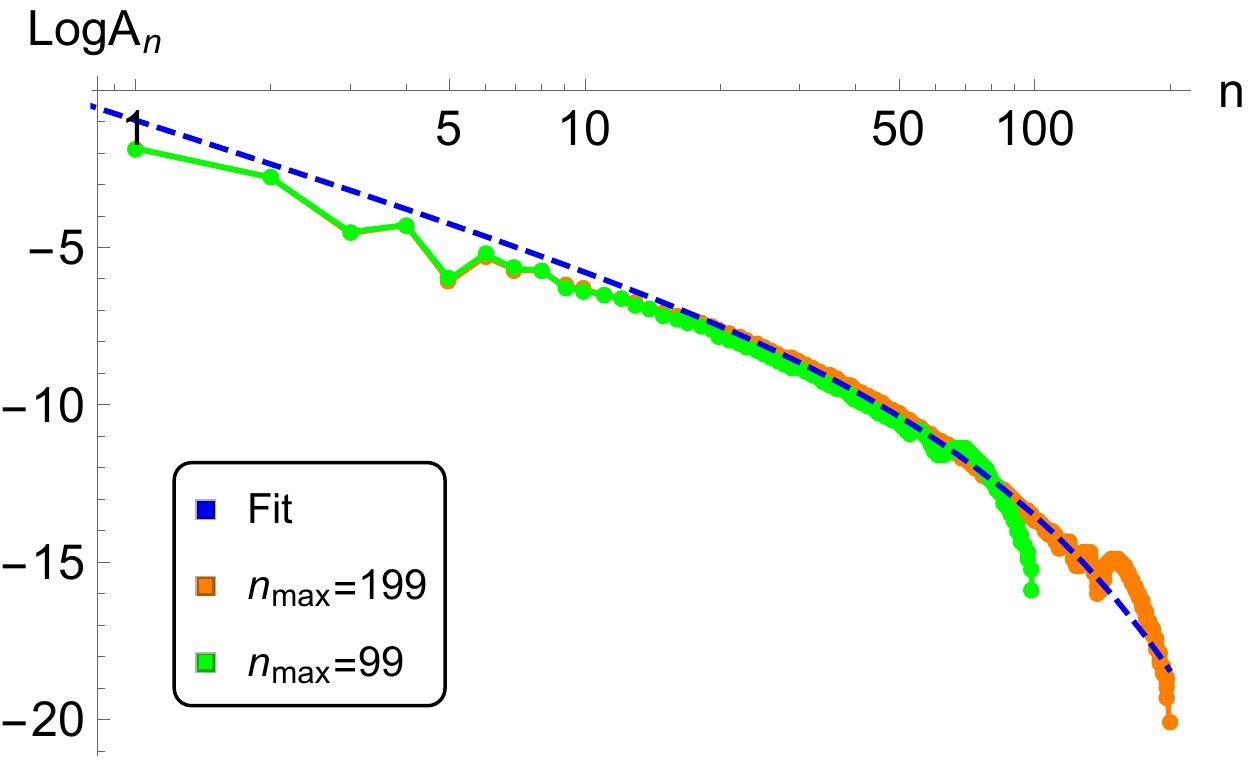}
\caption{{\bf Left:} Fitting of the amplitude spectrum at $\tau=\tau_{\text{max}}\sim0.485$ according to the formula (\ref{eq:ASansatz}) {\bf Right:} Fitting of the amplitude spectrum at $\tau=1$. At this time we find good agreement up to a maximum mode $n\sim50$ for $n_{\text{max}}=99$, and $n\sim100$ for $n_{\text{max}}=199$. The numerical noise observed for higher modes is due to the artificial cutoff, and originates because the energy starts to flow back into lower modes once $n_{\text{max}}$ is reached.
\label{FitEEtmax}}
\end{center}
\end{figure}

In Fig.~(\ref{FitEEtmax}) we plot the fitting function (\ref{eq:ASansatz}) at $\tau=\tau_{\text{max}}=0.485$, finding an excellent agreement even outside of the range of $n$ that we considered for the fits. This suggests that the formula (\ref{eq:ASansatz}) is actually quite robust, at least for the times at which all the parameters of the fit $\{\alpha(\tau),\gamma(\tau),\rho(\tau)\}$ are numerically accurate, i.e. for $\tau\leq\tau_{\text{max}}$. We also plot the fitting function (\ref{eq:ASansatz}) at $\tau=1$, finding good agreement in an intermediate range of modes, $n\in(20,50)$ for the $n_{\text{max}}=99$ truncation and $n\in(20,100)$ for the $n_{\text{max}}=199$. The numerical noise that that appears at higher modes is due to the artificial cutoff, and originates because the energy starts to flow back into lower modes once the initial cascade reaches $n_{\text{max}}$. This effect is delayed as we increase the number $n_{\text{max}}$, hence increasing the time of validity of the solution.

\begin{figure}[tb]
\begin{center}
\includegraphics[width=7.4cm]{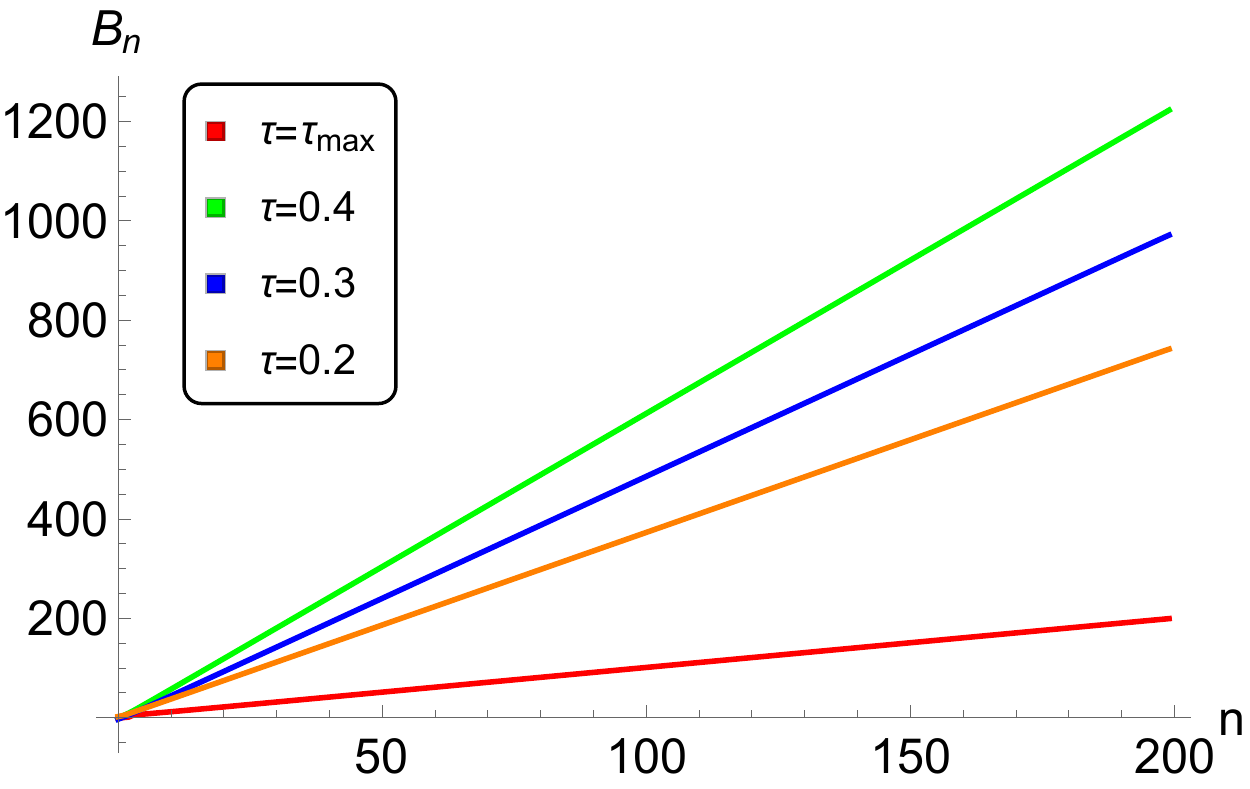}
\includegraphics[width=7.4cm]{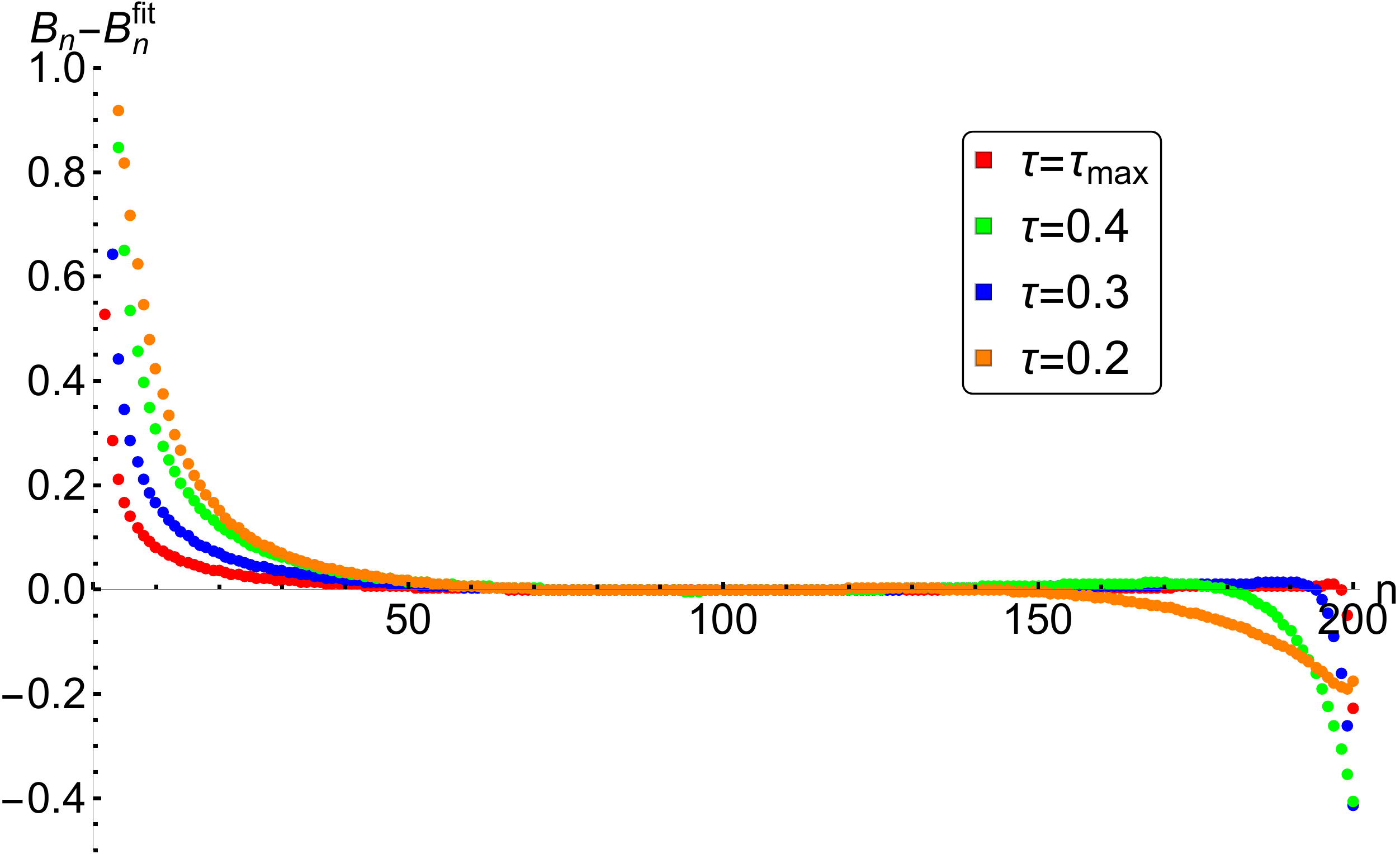}
\caption{ {\bf Left:} Unwrapped phases $B_n$ as a function of $n$ for various values of $\tau$. In all cases we see that the phases are perfectly aligned to form a straight line as in (\ref{eq:phasecoherence}). {\bf Right:} The difference between the fitted and the actual values as a function of $n$.
\label{fig:phasesEE}}
\end{center}
\end{figure}

It is also interesting to test the validity of the coherent phase conjecture \cite{FreYan15} directly with the phases $B_n(\tau)$ of our numerical evolution, even though we do not observe collapse. In order to do this, we plot in Fig.~(\ref{fig:phasesEE}) $B_n$ as a function of $n$ for various values of $\tau$.\footnote{Since $\text{Arg}[\alpha_n]=B_n\in(-\pi,\pi)$ we first \emph{unwrap} the phases before plotting the actual values. This amounts to shift $B_n\to B_n+2\pi k$, $k\in\mathbb{Z}$ every time that $\Delta B_n=B_{n+1}-B_{n}$ changes sign.} Interestingly, we see an almost perfect line for $\tau\leq\tau_{\text{max}}$ (recall that the inverse cascade kicks in approximately at this time), with deviations from linearity being indistinguishable to the naked eye. To quantify the linearity of the spectrum we subtract the fitted values from the actual data. The results are presented in Fig.~(\ref{fig:phasesEE}) as well (right panel). We observe a very good agreement with a linear spectrum, especially in the range of values used for the fit, $60 \leq n < 140$. As mentioned before, this can also be seen from the right panel of Fig.~(\ref{fig-AnEqualE}), specifically from the comparison of the upper envelope of $\Pi(x=0)$ and the red curve (an upper bound that assumes perfect phase alignment). We see that at $\tau=\tau_{\text{max}}$ and at each peak of the different bounces the perfect phase alignment is in good approximation with the actual upper envelope. Conversely, at each valley of the bounces the approximation worsens. These two observations combined suggest that energy transfer via direct cascades improves the coherence of the spectrum, while inverse cascades tend to decohere the phases. It would be interesting to quantify the interplay between the two in more detail, but we leave it for future studies.

$\quad$\\
\noindent \underline{{\bf Non-equal energy:}}\\

Other values of $\kappa$ in (\ref{eq-An2M}) imply that the initial energy of the two modes differ. We evolved the system for a wide range of $\kappa$ and repeated the analysis presented for the case $\kappa=3/5$. In extreme cases where most of the energy is deposited in one of the modes, the system can be though of a small perturbation of a single mode solution, which is known to belong to one of the stability islands \cite{BucGre15,GreMai15,DimYan15}. In such cases we do not expect black hole formation. On the other hand, values of $\kappa$ close to $\kappa=3/5$ should behave similarly to the equal energy data and are expected to ``collapse'' at infinite $\tau$.

\begin{figure}[t!]
\begin{center}
\includegraphics[width=7.9cm]{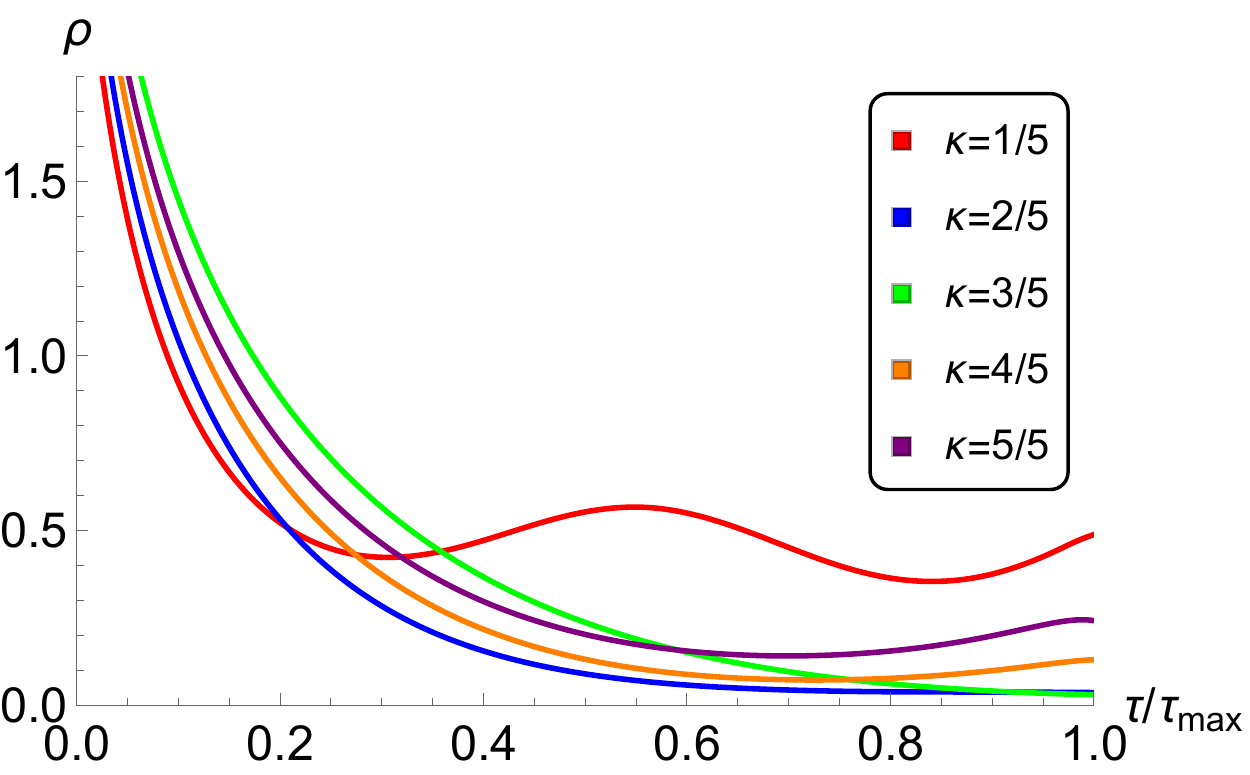}
\caption{The evolution of the analyticity strip width $\rho(\tau)$, for different ratios of the energy in the first two modes.
\label{fig-rhoNEE}}
\end{center}
\end{figure}

In Fig.~(\ref{fig-rhoNEE}) we plot the analyticity strip width $\rho(\tau)$ as a function of $\tau$ for various values of $\kappa$, showing different kind of behaviors.
For $\kappa=\{1/5,4/5,5/5\}$ $\rho(\tau)$ develops oscillations and never reach zero. These initial conditions are not exactly periodic as the single--mode solutions but exhibit revivals at time scales of order $1/\epsilon^2$ \cite{AbadaS14,daSilva:2014zva}. For $\kappa=2/5$ we do not see oscillations before $\tau_{\text{max}}$ but by doing the fit we find that $\rho_\infty$ is marginally above $1/n_{\text{max}}$. We conjecture that this case is close to the borderline of a subspace of solutions that collapse at infinite $\tau$, which for the two-mode initial data (\ref{eq-An2M}) is given by an open set with $\kappa\in(\kappa_{\text{min}},\kappa_{\text{max}})$. We further studied the phase spectrum of all these initial data, and found always good agreement with the coherent phase ansatz up to the time $\tau_{\text{max}}$.

\subsection{A speed limit for energy transfer\label{sec:speed}}

It is interesting and surprising that some initial data appears to cascade to arbitrarily high frequencies in finite time, while other initial conditions take an infinite time. To really convince ourselves that the energy cascade takes infinite slow time for some initial conditions, we need an analytic argument. In this section, we give a simple argument showing that for solutions where two low modes dominate the spectrum, the remaining modes obey a simple speed limit in how fast energy can propagate to high frequencies. This speed limit depends on frequency, and has the property that in 3+1 dimensions it takes infinite slow time to reach infinity, while in higher dimensions it takes finite time.

Our analysis rests on the strong assumption that the solution is dominated by two low modes, and it would be very interesting to relax this assumption. Despite its limitations, we find our analytic argument worthwhile because it agrees with existing numerics, and it may point the way towards a more rigorous analytic argument.

 \begin{figure}[tb]
\begin{center}
\includegraphics[width=7.9cm]{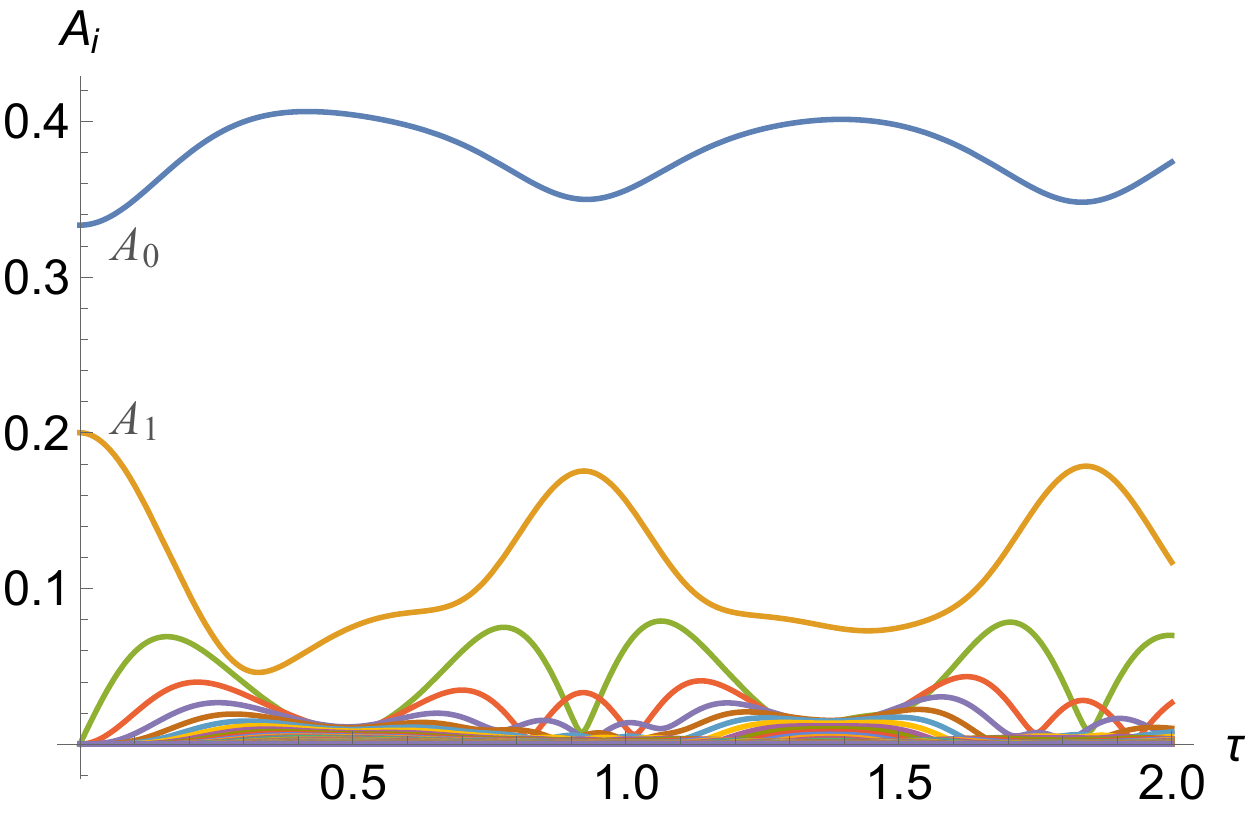}
\caption{Evolution of amplitudes for the equal-energy two-mode data using a $n_{\text{max}}=199$ mode truncation. From this plot it is clear that the modes $A_0$ and $A_1$ dominate throughout the evolution, at least up to $\tau=2$. We recall that already at this time there have been two bounces in the Ricci scalar (see right panel of Fig.~(\ref{fig-AnEqualE})), and both direct and inverse cascades are present after $\tau>\tau_{\text{max}}\sim0.485$.\label{fig:amplitudes}}
\end{center}
\end{figure}

In Fig.~(\ref{fig:amplitudes}) we show the evolution of amplitudes for two-mode initial conditions with equal energy. As seen form this figure, the lower two modes dominate the evolution for the whole time of the numerical simulation. We will assume this behavior extrapolates to later times.\footnote{Notice that this assumption \emph{does not} exclude the existence of a power-law spectrum: the solution can be dominated by the first 2 modes and still have a power law at large mode number.} For this purpose, it is convenient to write the equations of motion for the truncated resonant system in terms of complex amplitudes $\alpha_n \equiv A_n \exp(i B_n)$. The equations of motion become
\begin{equation}
2 i \omega_n {d \alpha_n \over d \tau} = \sum_{i+j = k + n} C_{ijkn} \alpha_i \alpha_j \bar{\alpha}_k
\end{equation}
Now we will assume that the solution is dominated by two low modes; for definiteness, take modes $n=0,1$, but the basic structure of our results will remain unchanged. In this case, we can keep only terms in the sum where two of the indices take the values $0, 1$. Since we have to obey the resonance condition $i+j = k+n$, for $n>2$ this means that $k = 0, 1$, leaving the equation
\begin{equation}
2 i \omega_n {d \alpha_n \over d \tau} = 2 C_{1 (n-1) 0 n}\ \alpha_1 \alpha_{n-1} \bar\alpha_0 + 2 C_{0 n 0 n} \alpha_0 \alpha_{n} \bar\alpha_0 + 2 C_{1 n 1 n}\alpha_1 \alpha_{n} \bar\alpha_1 + 2 C_{0 (n+1) 1 n}\alpha_0\alpha_{n+1} \bar\alpha_1
\end{equation}
Collecting terms and using the symmetries of the $C$ coefficients, we get
\begin{equation}
 i \omega_n {d \alpha_n \over d \tau} - \left( C_{0 n 0 n} \alpha_0 \bar\alpha_0  +  C_{1 n 1 n}\alpha_1 \bar\alpha_1\right) \alpha_{n} = C_{1 (n-1) 0 n}\ \alpha_1  \bar\alpha_0 \alpha_{n-1} + C_{1 n 0 (n+1) }\alpha_0 \bar\alpha_1 \alpha_{n+1}
 \end{equation}
Despite the complicated coefficients, this equation has two simplifying features:
\begin{itemize}
\item It is linear in the amplitudes $\alpha_n$.
\item It is {\it local}: a given mode is only influenced by its nearest neighbors (in addition to modes 0 and 1, which we think of as a background.)
\end{itemize}

We are interested in large mode numbers $n$. Since the $C$ coefficients are smooth functions, and $n$ is large, we take $C_{1 n 0 (n+1)} \approx C_{1 (n-1) 0 n} \equiv C_n$. We treat the dominant modes $\alpha_{0,1}$ as constants (it would be interesting to allow for time dependence).

As discussed above, the overall phase and the relative phase between any $two$ modes is unphysical, so we are free to choose convenient phases for $\alpha_{0,1}$. We use this freedom to take the product $\alpha_0 \bar\alpha_1$ to be pure imaginary for convenience, leading to the equation
\begin{equation}
 i \omega_n {d \alpha_n \over d \tau} - \left( C_{0 n 0 n}A_0^2  +  C_{1 n 1 n}A_1^2\right) \alpha_n = i C_{1 (n-1) 0 n}\ A_1 A_0 (\alpha_{n-1} - \alpha_{n+1})
 \end{equation}
 This equation can be written in the simpler notation
 \begin{equation}
 {d \alpha_n \over d \tau} + i d_n \alpha_n  + {1 \over 2} c_n (\alpha_{n+1} - \alpha_{n-1}) = 0
 \end{equation}
 with the definitions
 \begin{equation}
 d_n \equiv {C_{0 n 0 n}A_0^2  +  C_{1 n 1 n}A_1^2 \over \omega_n} \ \ \ \ \ \ \ \ \ \ \ c_n \equiv {2 C_{1 (n-1) 0 n}\ A_1 A_0 \over \omega_n}
 \end{equation}
 Since a given mode couples to nearest neighbors, we would like to identify the speed at which signals can propagate in frequency space. To estimate this, take a continuum limit of the above equation, treating $n$ as a continuous variable, to get
 \begin{equation}
  {\partial \alpha \over \partial \tau} + i d_n \alpha  +  c_n {\partial \alpha \over \partial n}= 0
  \end{equation}
  We have not fully analyzed this equation, but the speed of propagation can be read off by comparing the coefficient of the time derivative to the space derivative, yielding a speed of propagation that depends on the mode number
  \begin{equation}
 {\rm speed} = {d n \over d \tau} = c_n
 \end{equation}
 It is now interesting to ask whether energy can propagate to arbitrarily large mode numbers in finite time. The time to reach infinitely large mode numbers is
 \begin{equation}
 \Delta \tau = \int^{\infty} {dn \over c_n}
 \end{equation}
 Whether this is infinite depends on the function $c_n$, which depends on the dimension. In all dimensions, $\omega_n \sim n$. In 3+1 dimensions, the interaction coefficient $C_{1 (n-1) 0 n} \sim n^2$ \cite{CraEvn15a, CraEvn15b}. Therefore, in 3+1 dimensions, $c_n \sim n$, and the integral is logarithmically divergent, and the cascade cannot reach infinity in finite time. The maximum excited mode number as a function of time can increase at most as $n_{\rm max} \sim \exp(a \tau)$. This agrees well with the observation in our numerical evolution that the spectrum has an exponential form $\exp(-\rho n)$ with $\rho \sim \exp(-a \tau)$.

 In higher than 3+1 bulk dimensions, we expect the interaction coefficients to scale with larger powers of $n$, leading to a scaling $c_n \sim n^p$ with $p > 1$. This renders the integral convergent, meaning that energy can reach infinite mode number in finite time. This is in good agreement with the results obtained in \cite{BizMal15} in 4+1 dimensions, where the spectrum was observed to approach a power law in finite time.

  As a reminder, we were able to reach this strong conclusion by making a strong assumption that modes 0 and 1 dominate throughout the evolution. We expect a similar result whenever we assume that the evolution is dominated by a finite number of low-frequency modes: the equation for the high modes will still become linear and quasi-local, with the degree of nonlocality in frequency space determined by the frequency of the low modes that are excited.

  On the other hand,  the assumption that the solution is dominated by a few low modes can clearly break down as energy is transferred to higher modes. For this reason, it is not at all clear that the `speed limit' found under this assumption is a robust result, although it is intriguing that it appears to agree with numerical results. It would be very interesting to derive a more generally valid speed limit on the rate of energy transfer.


\section{Gaussian Data}\label{sec:GaussianData}

In this section we study the evolution of scalar fields with Gaussian initial data, ${\phi (0,x )=2 \text{exp}(-\frac{\text{tan}^2 x}{\sigma ^2})}$. Previous simulations in the full GR system reported collapse for sufficiently narrow profiles \cite{BucLie13,MalRos13b}, namely for $\sigma < 0.3$. We present results for two cases, $\sigma=0.15$ and $\sigma = 0.25$. We follow the same analysis as in Sec.(\ref{sec:2ModeData}), so we will not repeat the details here. However, it is worth mentioning that this case is considerable simpler than the two-mode data, since these initial conditions collapse at finite slow time $\tau$. We find that the time at which the analyticity strip becomes zero, $\tau=\tau _{\star}$, is not very sensitive to the specific $n_{\text{max}}$ that we use (as long as it is sufficiently high), so we will only show results for $n_{\text{max}}=199$. In this case we have $\tau _{\text{max}} = \tau _{\star}$, namely, we trust the evolution up to the time of the collapse.

\begin{figure}[tb]
\begin{center}
\includegraphics[width=7.4cm]{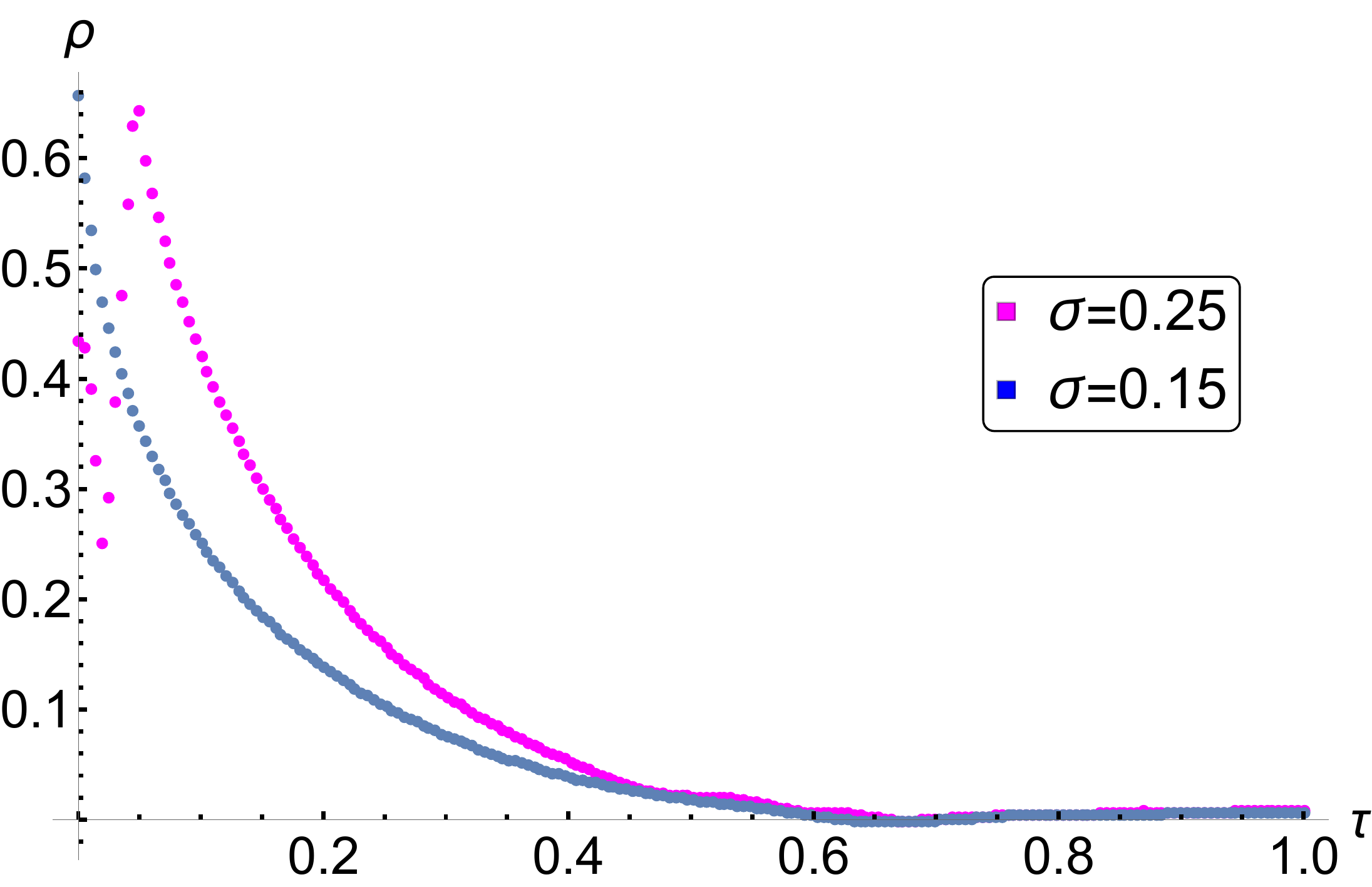}
\caption{ The evolution of the analyticity strip width $\rho (\tau)$ for Gaussian data with $\sigma=0.25$ and $\sigma=0.15$. In both case $\rho$ becomes zero in finite slow time $\tau _{\star} \sim 0.625$.
\label{fig-GaussianAnalyticityStrip}}
\end{center}
\end{figure}

In Fig~(\ref{fig-GaussianAnalyticityStrip}) we plot the evolution of the analyticity strip width, for the two above-mentioned data to make precise the contrast with the 2--mode data. Here $\rho (\tau)$ goes to zero in {\bf finite slow time} $\tau_{\star} \sim 0.625$. Our results confirm the observation of \cite{BucLie13}, namely that narrow profiles lead generically to black hole collapse. However, we observe a small deviation with respect to \cite{MalRos13} regarding the resulting power-law of the energy cascade. More specifically, we find that at the time of collapse, i.e. when the analyticity strip width $\rho (\tau)$ goes to zero, the power $\gamma (\tau)$ approaches the value $3/2$ instead of $8/5$. Notice that the exponent $3/2$ is precisely the value predicted by the coherent phase ansatz \cite{FreYan15}.

\begin{figure}[tb]
\begin{center}
\includegraphics[width=7.4cm]{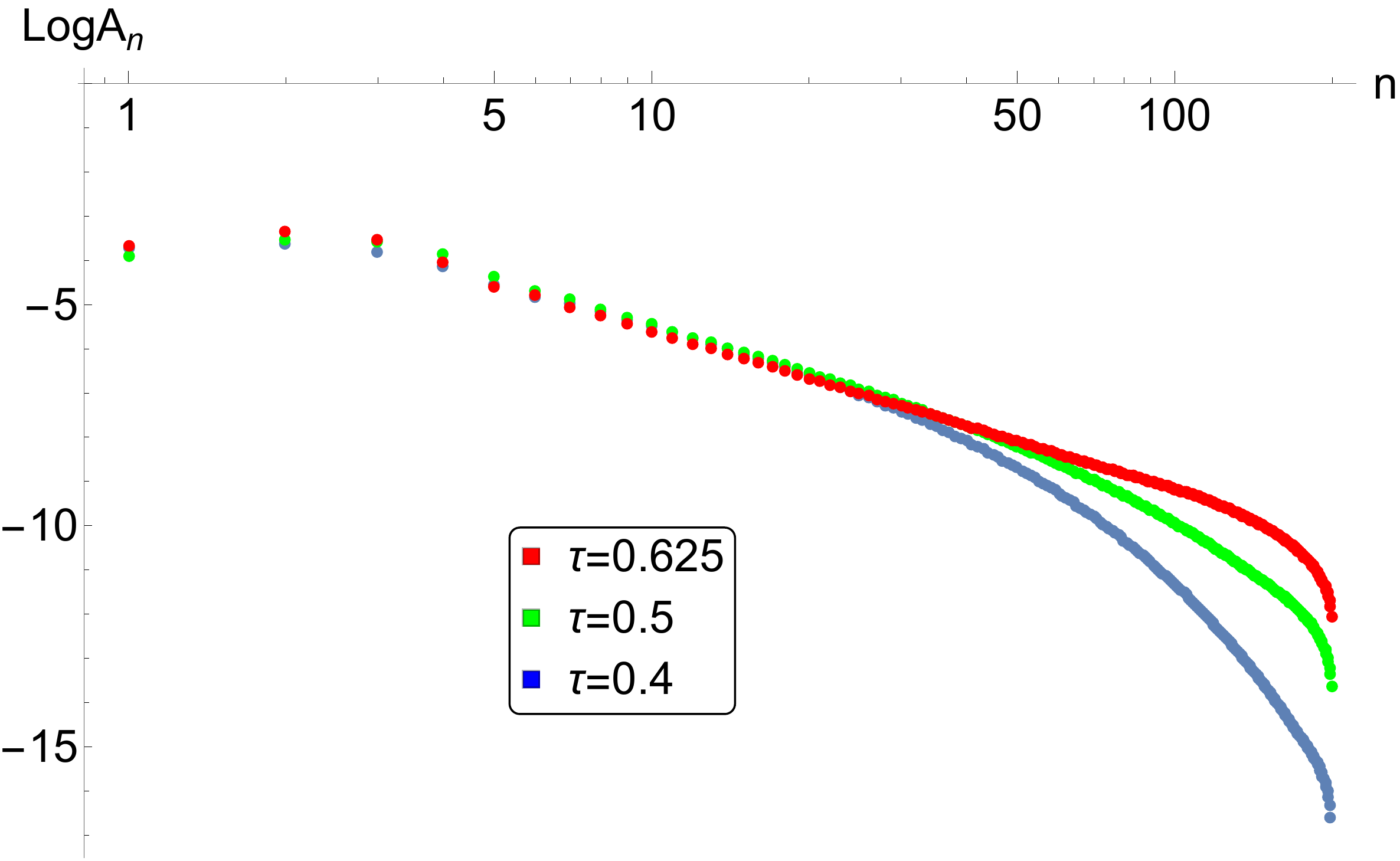}
\includegraphics[width=7.4cm]{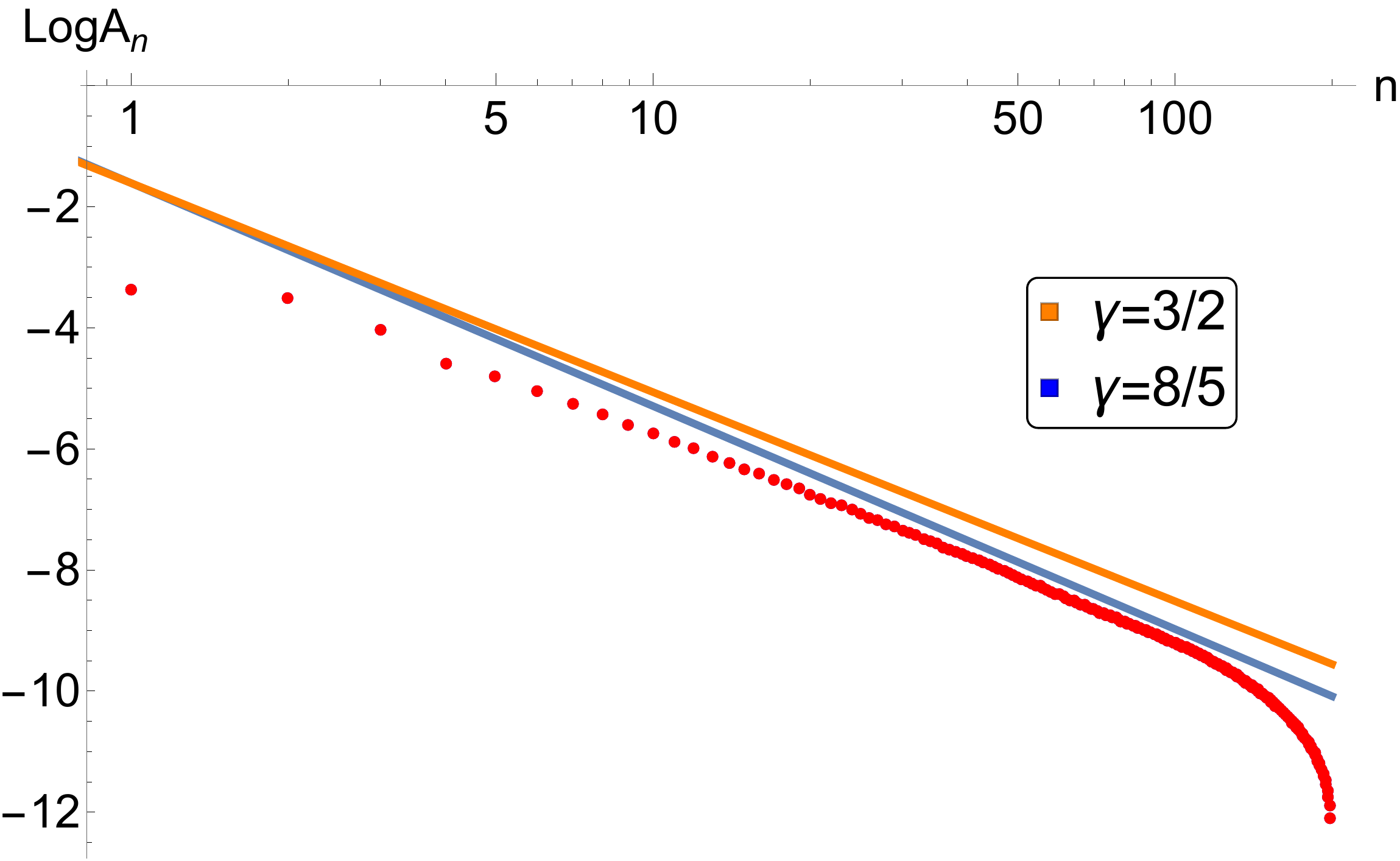}
\caption{ {\bf Left}: The evolution of the amplitude spectra for Gaussian data with $\sigma=0.15$ towards a power law $\gamma (0.625) \sim 3/2  $ in the TTF system. {\bf Right}: Comparison of the two power-laws with the actual data at the time of \emph{collapse}, $\tau _{\star} = 0.625$.
\label{fig-PowerLawGaussian15}}
\end{center}
\end{figure}

\begin{figure}[tb]
\begin{center}
\includegraphics[width=7.4cm]{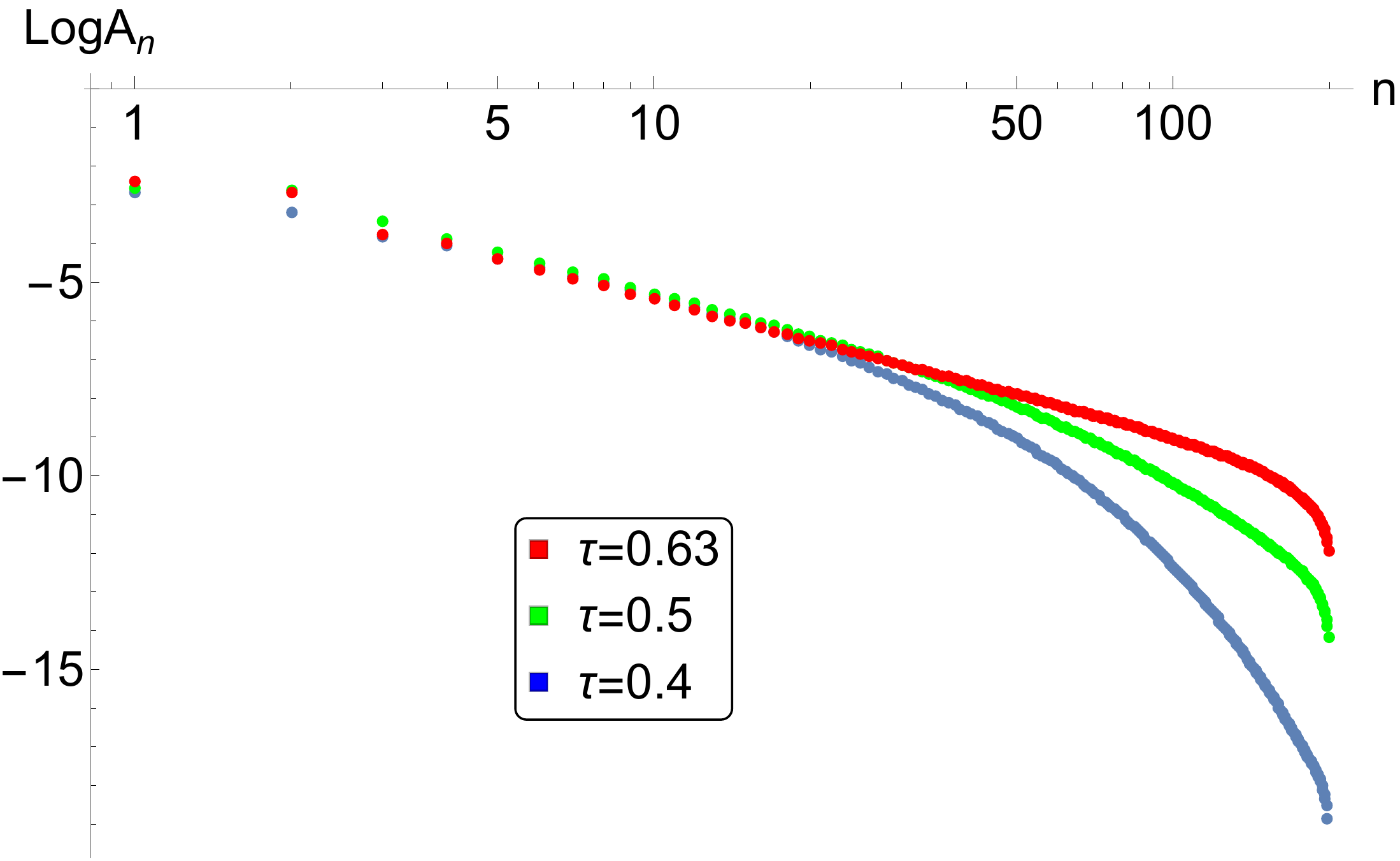}
\includegraphics[width=7.4cm]{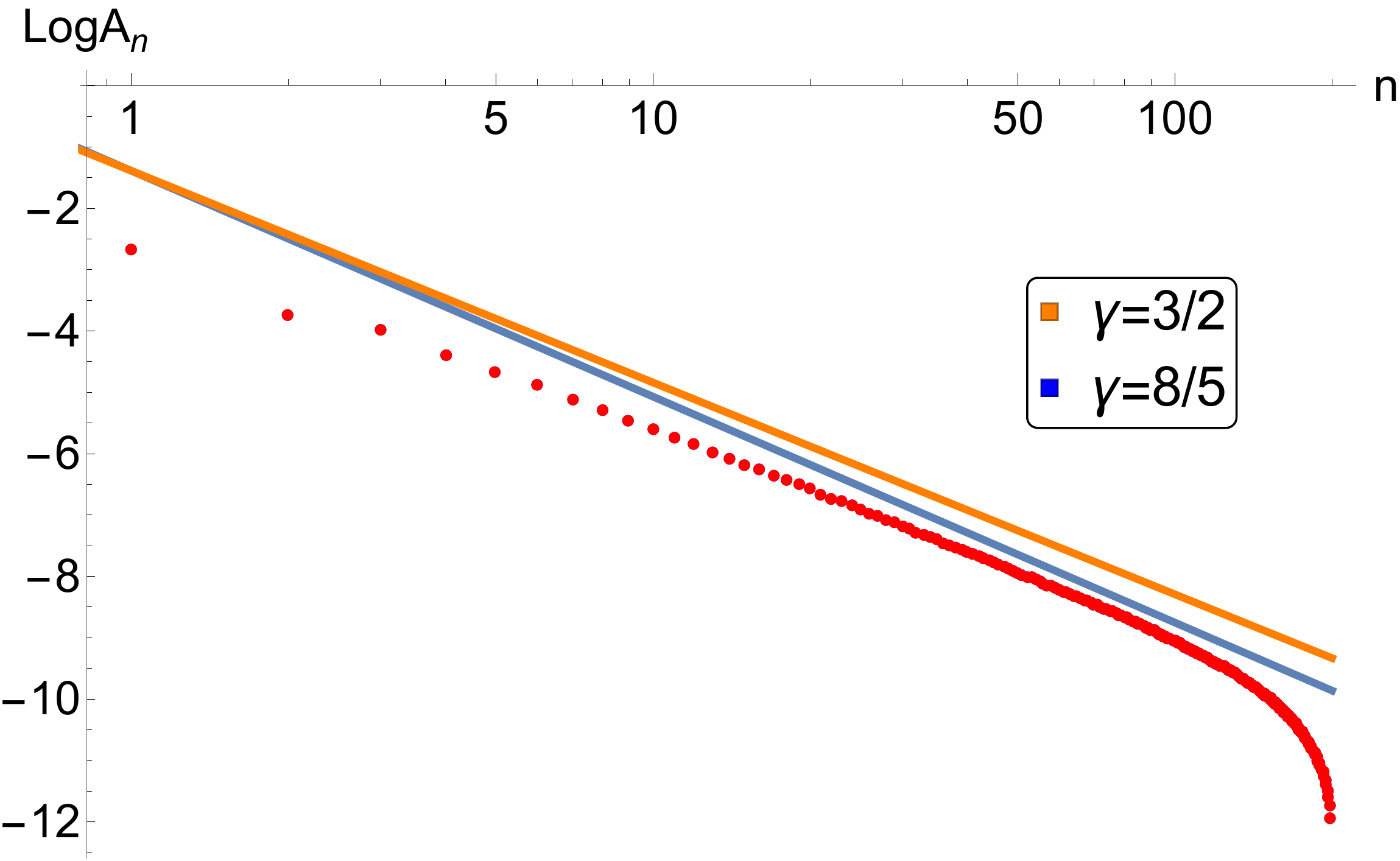}
\caption{ {\bf Left}: The evolution of the amplitude spectra for Gaussian data with $\sigma=0.25$ towards a power law $\gamma (0.63) \sim 3/2  $ in the TTF system. The evolution is almost identical to the case of $\sigma = 0.15$. {\bf Right}: Comparison of the two power-laws with the actual data at the time of \emph{collapse}, $\tau _{\star} \sim 0.63$.
\label{fig-PowerLawGaussian25}}
\end{center}
\end{figure}

In Fig.(\ref{fig-PowerLawGaussian15}) and Fig.(\ref{fig-PowerLawGaussian25}) we present the evolution of the amplitude spectra for Gaussian initial data with $\sigma = 0.15$ and $\sigma = 0.25$ respectively, towards a power law and we contrast the two values of $\gamma$ mentioned above. We see that indeed the value $\gamma (\tau _{\star})=3/2$ is a better fit. However, we notice that close to the time of collapse $\tau _{\star}\approx0.625$ the numerics are subtle and, although our results strongly suggest a power law of 3/2, another value for $\gamma$ very close to this one is still a possibility.

To illustrate the subtleties of the numerical methods we now study into more detail the fitting methods for the case of the $\sigma = 0.15$ Gaussian data close to the \emph{collapse} point $\tau_{\star} \sim 0.625$. We fit the data in the range $30 \leq n \leq 110$ in order to to take into account only the asymptotic form of the spectrum and avoid cut-off effects. In Table~\ref{table-fits}  we present the fitting functions for three different values of $\tau = \{ 0.62 , 0.625, 0.63 \}$ and we fit both a power law spectrum and a power law spectrum with an exponential decay. We observe that the analyticity strip width turns from positive, at $\tau = 0.625$ to negative, at $\tau = 0.63$ which means that it hits zero in finite slow time. Even though the value of the exponent $\rho$ is very small close to the moment of collapse, our fits show that it plays a significant role in estimating the exact value of the power $\gamma$. Moreover, the exact range of data that we use for the fitting also affects the result for $\gamma$.\footnote{We chose the above-mentioned range, $n \in [30,110]$, since it represents the most linear part of the plot and neglects the cut-off effects.} The fits, in both cases suggest a power law very close to 3/2, however as we mentioned earlier, values close to this one are not completely ruled out. This happens because the convergence of $\gamma$ is not sufficiently fast for times approaching the collapse.

\begin{table*}\scriptsize \centering
\ra{1.3}
\begin{tabular}{@{}rrrrcrrrcrrr@{}}\toprule
& \multicolumn{2}{c}{$\tau = 0.62$} & \phantom{ab}& \multicolumn{2}{c}{$\tau=0.625$} &
\phantom{ab} & \multicolumn{2}{c}{$\tau=0.63$}\\
\cmidrule{2-3} \cmidrule{5-6} \cmidrule{8-9}
& \emph{Parameter } & \emph{Error}  &&  \emph{Parameter } & \emph{Error} &&  \emph{Parameter } & \emph{Error}  \\ \midrule
$\mathbf{A_n \sim \alpha n^{-\gamma} e^{-\rho n}}$\\
$\text{Log}(\alpha)$ & -2.17944 & 0.02399  && -2.18595 & 0.0301 && -2.20568 & 0.03769 \\
$\gamma$ & 1.51033 &  0.0077 && 1.51579 & 0.00966 && 1.517 & 0.012 \\
$\rho$ & 0.000754 & 0.00012 && 0.000205 & 0.000151 && -0.000272 & 0.000189 \\
$\mathbf{A_n \sim \alpha n^{-\gamma} } $\\
$\text{Log}(\alpha)$ & -2.03288 & 0.00649 && -2.14604 & 0.00672 && -2.25847 & 0.00843  \\
$\gamma$ & 1.55796 & 0.00154 && 1.52876 & 0.0016 && 1.49985 & 0.002 \\
\bottomrule
\end{tabular}
\caption{The fitting values and the corresponding errors for three different times very close to the collapse time $\tau_{\star} \sim 0.625$, in the case of a power law with an exponential ({\bf up}) and a power law alone ({\bf down}).}
\label{table-fits}
\end{table*}

\begin{figure}[tb]
\begin{center}
\includegraphics[width=7.4cm]{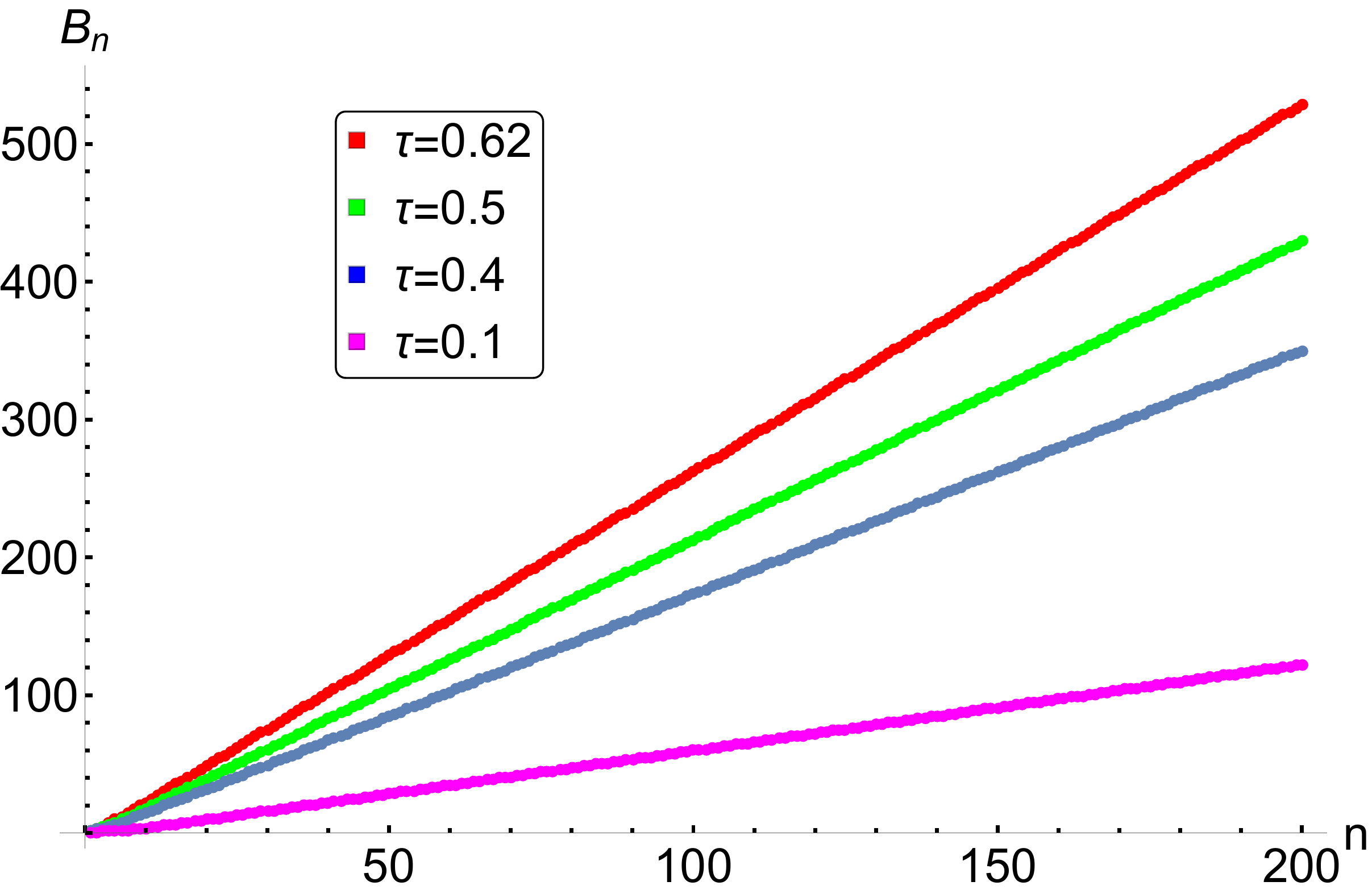}
\includegraphics[width=7.4cm]{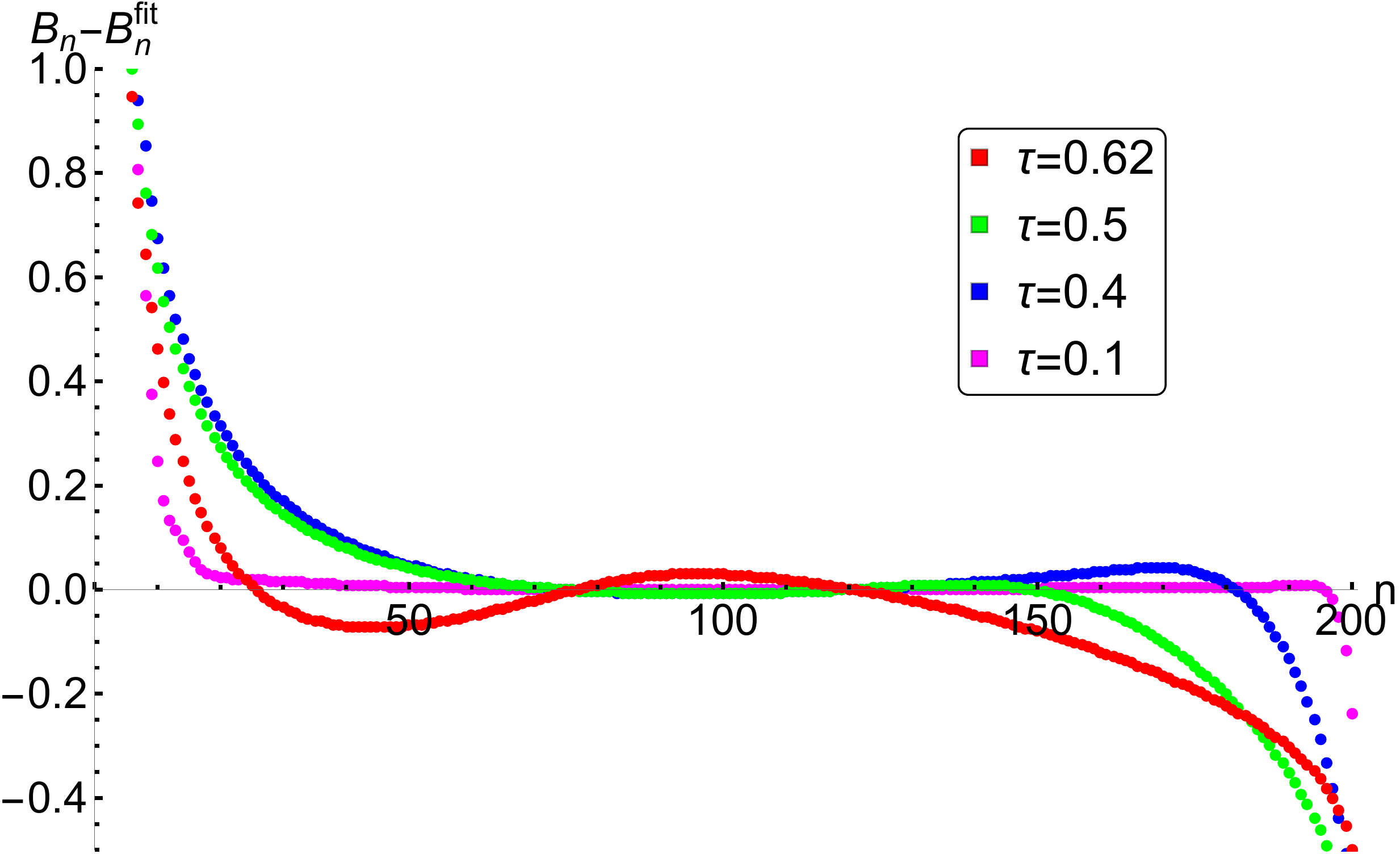}
\caption{ {\bf Left:} The phases of the modes as a function of the mode number $n$ when \emph{unwraped} for different slow times $\tau$. {\bf Right:} The difference between the fitted and actual values as a function of $n$.
\label{fig-PhasesGaussian}}
\end{center}
\end{figure}

In Fig.(\ref{fig-PhasesGaussian}) we present again the phase-spectrum as a function of the mode number $n$ for different slow times. As we did earlier, we fit the data to a linear function and we quantify the deviation from linearity by subtracting the fitting value from the actual data. We see once more that the linear fit is a very good approximation, however interesting patterns appear, especially for late times, that might suggest towards a slight improvement to the perfectly coherent spectrum. We hope we will come back to this issue in a future work.

\section{Discussion and outlook}

In this work we presented numerical evidence and argued that the turbulent cascade of energy towards modes of higher frequency in the problem of the instability of global AdS happens in a \emph{phase--coherent} way, in stark contrast with the standard theory of weak turbulence. However, our results leave a small room for possible deviations on exact coherence, which we believe is something interesting to look at in the future. It is worth noting that the perturbations that we studied here are initially phase coherent so one might think that this could play a role in the later-time coherence of the phases. A natural generalization would be to study the development of phase coherence in perturbations that are not initially phase coherent.\footnote{We will come back to this issue in future work, but some preliminary results that we have suggest that even in data for which the phases are initially randomly distributed, the higher modes are excited coherently.}

In our numerical studies we have considered two kinds of initial conditions, the so-called two-mode data and Gaussian data. We have specific results for each case. Our results for the two-mode equal-energy data suggest that the width of the analyticity strip vanishes in the limit $\tau \rightarrow \infty$. This would imply that the time scale required to collapse is slightly larger than the standard one, e.g. $t\sim\epsilon^{-2}\log\epsilon$ instead of $t\sim\epsilon^{-2}$, so the collapse might not be fully captured within TTF. We observed that the same holds for small deviations of the equal-energy data so we conjectured that there is an open set of initial conditions leading to such \emph{slow cascades}. Moreover, we gave an analytic argument in support of this idea based on a speed limit for energy transfer and showed that it may happen in $3+1$ dimensions but not in higher dimensions. Needless to say, it would be interesting to revisit this case in the full GR system.

For Gaussian initial data with narrow profiles we observed that the spectrum of the amplitudes approaches a power-law at finite slow time $\tau=\tau_\star$. Our results suggest that the power at the time of the collapse is given by $\gamma (\tau _{\star}) \sim 3/2$, correcting a previous value reported in the literature. Interestingly, the power $3/2$ agrees with the value predicted by the coherent phase ansatz. It would be very interesting to determine the power-law in the case of the two mode data and see if it agrees with the above value, as well as to obtain a definite answer for the precise value in the case of the Gaussian data, perhaps through analytic techniques.

\acknowledgments

We thank I-Sheng Yang, Joanna Jalmuzna, Javier Mas, Oscar Dias, Jorge Santos, Andrzej Rostworowski, Ben Craps and Antonio Rotundo for useful discussions. This work is part of the $\Delta$-ITP consortium and also supported in part by the Foundation for Fundamental Research on Matter (FOM), both are parts of the Netherlands Organisation for Scientific Research (NWO) that is funded by the Dutch Ministry of Education, Culture and Science (OCW). FD is supported by GRAPPA PhD Fellowship. The research of JFP is supported by the Netherlands Organization for Scientific Research (NWO) under the VENI scheme.

\bibliographystyle{JHEP}
\bibliography{all_active}

\end{document}